\documentclass{pasj01}

\Received{$\langle$reception date$\rangle$}
\Accepted{$\langle$acception date$\rangle$}
\Published{$\langle$publication date$\rangle$}

\newcommand{\cmmnt}[1]{}

\usepackage{savesym}
\savesymbol{iint}
\savesymbol{iiint}
\savesymbol{iiiint}
\savesymbol{idotsint}
\savesymbol{leftroot}
\savesymbol{uproot}
\usepackage{amsmath}
\restoresymbol{AMS}{iint}
\restoresymbol{AMS}{iiint}
\restoresymbol{AMS}{iiiint}
\restoresymbol{AMS}{idotsint}
\restoresymbol{AMS}{leftroot}
\restoresymbol{AMS}{uproot}

\usepackage{amsmath} 
\usepackage{amstext}
\usepackage{color}
\usepackage{graphicx}
\usepackage[nointegrals]{wasysym}

\usepackage{url}

\usepackage{natbib}
\usepackage{tablefootnote}

\usepackage[title]{appendix}

\begin{document} 

\Received{2022/12/06}
\Accepted{2023/03/06}
\Published{}

\title{Detection of a new molecular cloud in the LHAASO J2108+5157 region supporting a hadronic PeVatron scenario\thanks{As section of the thesis to be submitted by Toledano--Ju\'arez as a partial fulfillment for the requirements of Ph. D. Degree in Physics, Doctorado en Ciencias (F\'isica), CUCEI, Universidad de Guadalajara}

}


\author{Eduardo \textsc{de la Fuente}\altaffilmark{1,2}
\thanks{Corresponding author: eduardo.delafuente@academicos.udg.mx}}

\author{Ivan \textsc{Toledano-Juarez}\altaffilmark{3}
}

\author{Kazumasa \textsc{Kawata}\altaffilmark{2}}

\author{Miguel A. \textsc{Trinidad}\altaffilmark{4}}

\author{Daniel \textsc{Tafoya}\altaffilmark{5}}

\author{Hidetoshi \textsc{Sano}\altaffilmark{6}}

\author{Kazuki \textsc{Tokuda}\altaffilmark{7}}

\author{Atsushi \textsc{Nishimura}\altaffilmark{8}}

\author{Toshikazu \textsc{Onishi}\altaffilmark{9}}

\altaffiltext{1}{Departamento de F\'{i}sica, CUCEI, Universidad de Guadalajara, Blvd. Marcelino Garc\'{i}a Barragan 1420, Ol\'{i}mpica, 44430, Guadalajara, Jalisco, M\'exico}
\email{eduardo.delafuente@academicos.udg.mx}

\altaffiltext{2}{Institute for Cosmic Ray Research, University of Tokyo, Kashiwa 277-8582, Japan}
\email{kawata@icrr.u-tokyo.ac.jp}

\altaffiltext{3}{Doctorado en Ciencias en F\'{i}sica, CUCEI, Universidad de Guadalajara, Blvd. Marcelino Garc\'{i}a Barragan 1420, Ol\'{i}mpica, 44430, Guadalajara, Jalisco, M\'exico}
\email{ivan.toledano9284@alumnos.udg.mx}

\altaffiltext{4}{Departamento de Astronom\'{i}a, Universidad de Guanajuato, Apartado Postal 144, 36000, Guanajuato, Guanajuato, M\'exico}
\email{trinidad@astro.ugto.mx}

\altaffiltext{5}{Department of Space, Earth, and Enviroment, Chalmers University of Technology, Onsala Space Observatory, 439 92 Onsala, Sweden}
\email{daniel.tafoya@chalmers.se}

\altaffiltext{6}{Faculty of Engineering, Gifu University, 1-1 Yanagido, Gifu 501-1193, Japan}
\email{hsano@gifu-u.ac.jp}

\altaffiltext{7}{Department of Earth and Planetary Sciences, Faculty of Science, Kyushu University, Nishi-ku, Fukuoka 819-0395, Japan}
\email{tokuda.kazuki.369@m.kyushu-u.ac.jp}

\altaffiltext{8}{Nobeyama Radio Observatory, National Astronomical Observatory of Japan (NAOJ), National Institutes of Natural Sciences (NINS), 462-2 Nobeyama, Minamimaki, Minamisaku, Nagano 384-1305, Japan}
\email{atsushi.nishimura@nao.ac.jp}

\altaffiltext{9}{Department of Physics, Graduate School of Science, Osaka Metropolitan University, 1-1 Gakuen-cho, Naka-ku, Sakai, Osaka 599-8531, Japan}
\email{tonishi@omu.ac.jp}

\author{Takashi \textsc{Sako}\altaffilmark{2}}
\email{sako@icrr.u-tokyo.ac.jp}

\author{Binita \textsc{Hona}\altaffilmark{10}}
\altaffiltext{10}{Department of Physics and Astronomy, University of Utah, 439 92 Utah, USA}
\email{binita.hona@utah.edu}

\author{Munehiro \textsc{Ohnishi}\altaffilmark{2}}  
\email{ohnishi@icrr.u-tokyo.ac.jp}

\author{Masato \textsc{Takita}\altaffilmark{2}}
\email{takita@icrr.u-tokyo.ac.jp}

\KeyWords{ISM: clouds --- ISM: molecules --- ISM: individual objects (LHAASO J2108+5157, Kronberger 82) --- cosmic rays --- gamma rays}

\maketitle

\begin{abstract}

PeVatrons are the most powerful naturally occurring particle accelerators in the Universe. The identification of counterparts associated to astrophysical objects such as dying massive stars, molecular gas, star-forming regions, and star clusters is essential to clarify the underlying nature of the PeV emission, i.e., hadronic or leptonic. We present $^{12,13}$CO(J=2$\rightarrow$1) observations made with the 1.85~m radio-telescope of the Osaka Prefecture University toward the Cygnus OB7 molecular cloud, which contains the PeVatron candidate LHAASO J2108+5157. We investigate the nature of the sub-PeV (gamma-ray) emission by studying the nucleon density determined from the content of HI and H$_2$, derived from the CO observations. In addition to MML[2017]4607, detected via the observations of the optically thick $^{12}$CO(J=1$\rightarrow$0) emission, we infer the presence of an optically thin molecular cloud, named [FKT-MC]2022, whose angular size is 1.1$\pm$0.2$^{\circ}$. We propose this cloud as a new candidate to produce the sub-PeV emission observed in LHAASO J2108+5157. Considering a distance of 1.7 kpc, we estimate a nucleon (HI+H$_2$) density of 37$\pm$14 cm$^{-3}$, and a total nucleon mass(HI+H$_2$) of 1.5$\pm$0.6$\times$10$^4$ M$_{\odot}$. On the other hand, we confirm that Kronberger 82 is a molecular clump with an angular size of 0.1$^{\circ}$, a nucleon density $\sim$ 10$^3$ cm$^{-3}$, and a mass $\sim$ 10$^3$ M$_{\odot}$. Although Kronberger 82 hosts the physical conditions to produce the observed emission of LHAASO J2108+5157, [FKT-MC]2022 is located closer to it, suggesting that the latter could be the one associated to the sub-PeV emission. Under this scenario, our results favour a hadronic origin for the emission.

\end{abstract}



\section{Introduction}
\label{sec:intro}

While human-manufactured particle accelerators can achieve relatively high energies, up to 14 TeV (in a collision) with the Large Hadron Collider, and is planned to reach values as high as 100 TeV via the future circular collider\footnote{https://fcc-cdr.web.cern.ch/} in the 2040 decade, it is known that in the Universe there are natural sources that can accelerate particles reaching energies up to hundreds of Exa eV (EeV). Thus, one of the fundamental questions in high energy physics is: what is the maximum energy to which nature accelerates particles in the Galaxy?. The answer to this question can be obtained by studying high-energy radiation and astroparticles (cosmic-ray, neutrinos, and gamma-rays) produced by one type of the most powerful natural particle accelerators: \textit{PeVatrons}. Although there is no consensus on its definition, a \textit{PeVatron} can be considered an astronomical object that can accelerate particles up to PeV energies.

Since high-energy particles produce gamma rays and X-rays, they are studied within the field known as Gamma-Ray Astrophysics (GRA). This new research field provides key knowledge to our understanding of the physical laws and mechanisms underlying particle acceleration. In addition, given that cosmic particle accelerators are also sources of non-thermal radiation, GRA creates a connection between the study of high-energy cosmic rays, neutrinos, and gravitational waves and the multi-wavelength astronomy. 

The energy regime of the emission studied in GRA can be divided in three: \textit{High Energy} (0.1-100 GeV), \textit{Very High Energy (VHE)} (0.1-100 TeV), and \textit{Ultra High Energy (UHE)} (0.1-100 PeV). Since its emergence as a research field, GRA has gone through three \textit{eras} sparked by major revolutions: 1.- The discovery of energetic phenomena in the Universe (e.g. the Galactic diffuse gamma rays and the gamma-ray pulsars; \textit{the GeV era}) by satellite gamma-ray detectors (e.g. EGRET, FERMI-LAT), 2.- The development of water Cherenkov (e.g. Milagro and HAWC) and imaging air Cherenkov observatories; the construction of extensive (surface) air shower arrays (e.g. MAGIC, H.E.S.S., VERITAS, ARGO-YBJ, Tibet AS$\gamma$), which discovered extremely powerful objects like supernova remnants, pulsar wind nebulae, TeV Halos objects, binaries systems, and active galactic nuclei); \textit{the TeV era}, and 3.- The rise of the UHE observations and the discovery of PeVatrons by HAWC, Tibet AS$\gamma$ and LHAASO observations: \textit{the PeV era}. 

While cosmic rays with energies $<$ 1 PeV are thought to have a Galactic origin, those with energies $>$ 1 EeV are thought to originate in extra-galactic sources. The transition in between is investigated through the study of \textit{PeVatrons}. Since 2019, the Tibet AS$\gamma$ \citep{Amenomori2019,Amenomori2021a,Amenomori2021b,Amenomori2021c}, HAWC \citep{Abeysekara2020,Abeysekara2021,Albert2020a}, and LHAASO \citep{Cao2021a,Cao2021b} experiments have detected a dozen of UHE sources on the Galactic plane. Thus, thanks to PeV astronomy, we became able to investigate particle acceleration within the Galaxy beyond the PeV energy regime. On the other hand, the question of whether only hadrons (protons) or leptons (electrons) should be considered in the acceleration process is critical to understand, e.g., the underlying nature of the observed gamma ray radiation. 

To explain the production of the UHE gamma rays, two scenarios have been proposed: 1. Neutral pions decay caused by the interaction of PeV protons with molecular material of the ambient interstellar medium (ISM), known as the \textit{hadronic scenario}, and 2. Inverse Compton (IC) scattering, caused by high-energy electrons scattering off ambient interstellar cosmic background photons, diffusive starlight and Galactic interstellar IR emission. This one is known as the \textit{leptonic scenario}. The detection and characterization of atomic hydrogen (HI) and molecular gas are crucial to estimate the density of nucleons that could be interacting with high-energy cosmic rays. This would shed light on determining which of the two proposed gamma-ray emission mechanisms is playing a major role. The most used proxy to probe the physical properties of the neutral molecular component of the ISM is the CO molecule. To this end, CO emission lines are widely used to trace the cold (10--30 K), and dense (10$^3$--10$^4$ cm$^{-3}$) molecular gas phase of the ISM in the Galaxy \citep[e.g][]{Sano2020,Fukui2021}. A detailed survey and study of CO emission from the ISM is necessary to characterize the physical conditions of molecular clouds and material lying nearby gamma-ray sources, which will allow to elucidate the nature and the origin of the high-energy cosmic rays. Therefore, without minimizing the co-acceleration (hadrons and leptons), carrying out a detailed study of the molecular gas in the vicinity of a \textit{PeVatron} is essential to explore the hadronic nature of the gamma ray emission.  

HAWC \citep{Abeysekara2021, Albert2021}, Tibet AS$\gamma$ \citep{Amenomori2021c} and LHAASO \citep{Cao2021a} have confirmed the presence of three PeVatrons candidates in the Cygnus constellation: 1.- The Cygnus Cocoon powered by Cygnus OB2 association \citep{Ackermann2011,Aharonian2019}, 2.- HAWC J2019+368 powered by PSR J2021+3651 and its associated pulsar wind \textit{Dragonfly} Nebula \citep{Albert2021}, and 3.- LHAASO J2108+5157 (J2108 hereafter) in Cygnus OB7 molecular cloud (COB7-MC hereafter). Within \citet{Cao2021a} catalogue, where all PeVatrons candidates to date are included, J2108 is probably the most intriguing one due to not having a clear or typical counterpart, namely a pulsar,  pulsar wind nebula (PWN in singular PWNe in plural), supernova remnant (SNR in singular, SNRs in plural) or a TeV Halo object \citep{Cao2021b}, albeit extended emission of CO is known to be present in the region. Observations of $^{12}$CO(J=1$\rightarrow$0), $^{13}$CO(J=1$\rightarrow$0) and C$^{18}$O(J=1$\rightarrow$0) line emission toward this source have been reported \citep{Dobashi1994,Dobashi2014}. In addition, prominent star-forming regions are found within a couple-degree size region centred at the PeV peak emission of J2108, e.g., Kronberger 80, Kronberger 82 (Kron 82 hereafter; e.g., \citealt{Kronberger2006}) and IRAS 21046+5110 (IRAS 21046 hereafter; e.g. \citealt{Kumar2006}).

In this work we present a pioneering study of the COB7 region where we compare $^{12}$CO(J=2$\rightarrow$1) and $^{13}$CO(J=2$\rightarrow$1) line emission observations with the PeV energy gamma-ray emission to investigate the origin and nature of the sub-PeV emission in J2108. In section~\ref{sec:over} we give an overview of the Cygnus region, J2108, and its vicinity. The observations are described in Section~\ref{sec:obs}. The methodology is explained in section~\ref{sec:methodology}. The results and discussion are presented in section~\ref{sec:results_discussion}. We conclude our work in section~\ref{sec:conclusion}.

\begin{figure}[!ht]
\begin{center}
\includegraphics[width=\columnwidth]{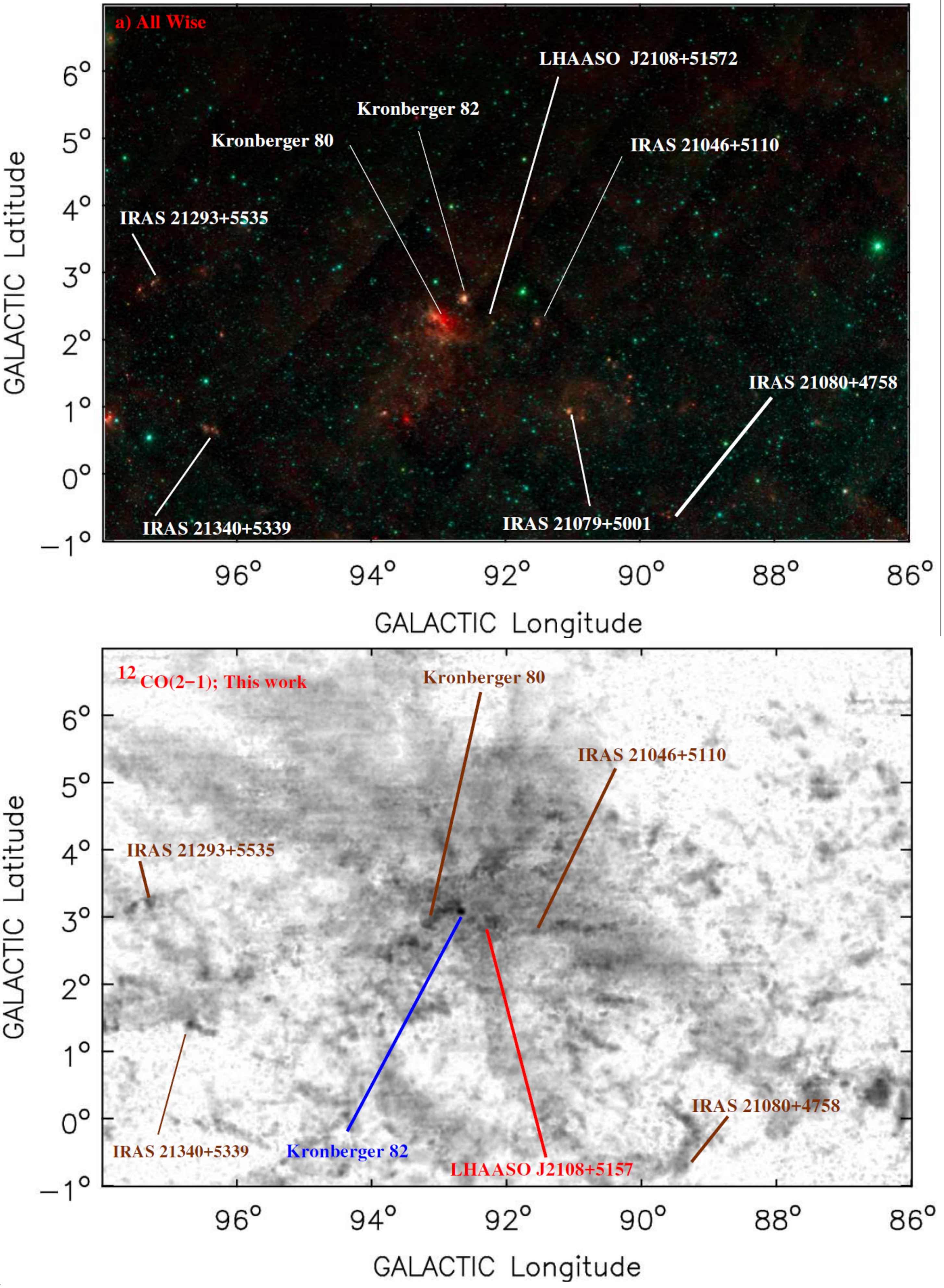}
\end{center}
\caption{\textbf{Top:} All WISE RGB (22$\mu$m +4.6$\mu$m +3.4$\mu$m) image of COB7-MC centred at l = 93$^{\circ}$ and b = +03$^{\circ}$. No clear counterpart is observed for J2108, although Kronberger 80, Kron 82, and IRAS 21046 are predominant in its neighbourhood. \textbf{Down:} The low resolution total flux integrated $^{\rm 12}$CO(2$\rightarrow$1) map \citep{Nishimura2020} discussed in the text. The brightest point corresponds to Kron 82 (blue marker). No clear condensation is observed on the J2108 position, but the vicinity is surrounded by molecular gas. Some IRAS sources and objects related to star-forming and star clusters are labelled for identification.}
\label{fig:CygOB7_intro}
\end{figure}

\begin{figure}[!ht]
\begin{center}

\includegraphics[width=\columnwidth]{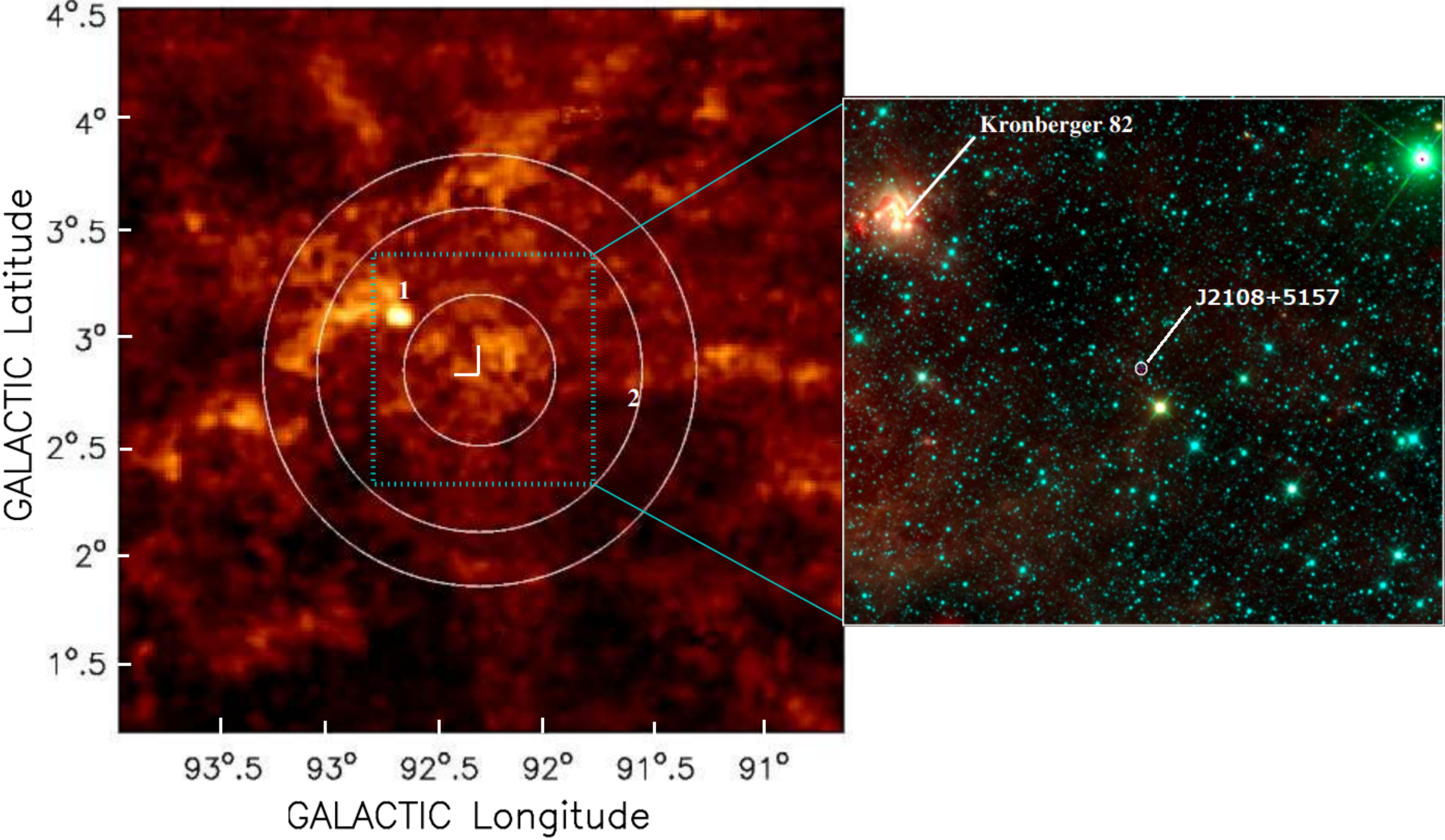}
\includegraphics[width=\columnwidth]{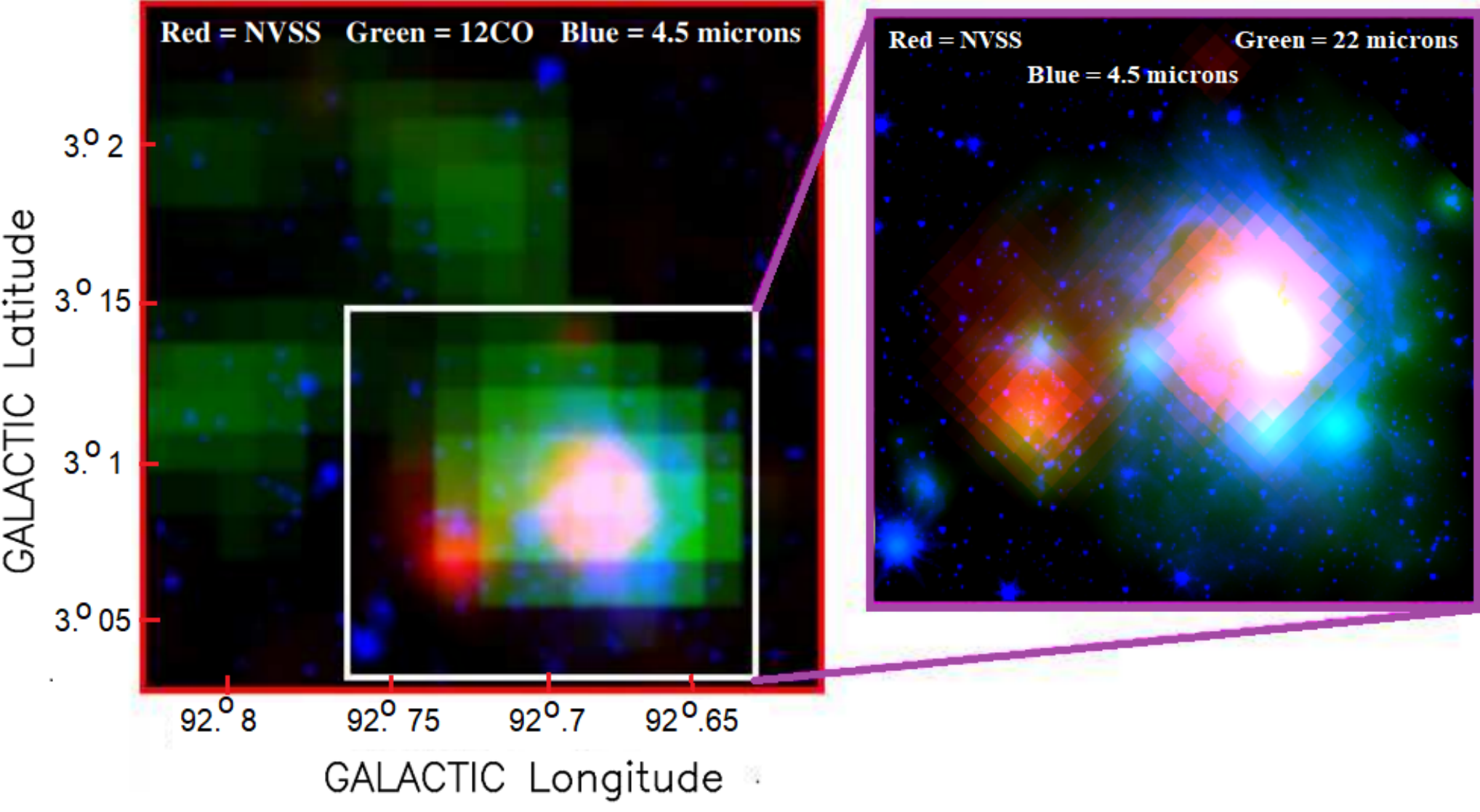}
\end{center}
\caption{\textbf{Top:} Same as Fig.~\ref{fig:CygOB7_intro}, but in a 3 square degree field and centred in J2108 (marks): $\alpha$(2000) = 21$^h$08$^m$36$^s$; $\delta$(2000) = +51$^{\circ}$57'00'' for the $^{\rm 12}$CO(2$\rightarrow$1) emission. The diameters of the circles are 0.7, 1.5, and 2$^{\circ}$ respectively. The regions related to Kron 82 (peak) and IRAS 21046+5110 are labelled as 1 and 2 respectively. All WISE RGB image (red = 22$\mu$m, green = 4.6$\mu$m, and blue = 3.4 $\mu$m) with a size of $\sim$ 1$^{\circ}$ is shown as an inset. The small circle shows the position of J2108. \textbf{Down:} A zoom of the zone labelled as 1 in the top left panel (Kron 82). This RGB image is NVSS (RC at 20 cm in red), our $^{12}$CO (2-1) data (in green), and the Spitzer--IRAC 4.5 $\mu$m image (in blue). Another RGB image (red = NVSS; green = WISE 22 $\mu$m; and blue = Spitzer--IRAC 4.5 $\mu$m) is shown as an inset. Thus, Kron 82 is the more prominent object across wavelengths. There are no RC or IR emissions in the vicinity of J2108 location.}
\label{fig:lhaaso_opt}
\end{figure}

\section{Sources Overview}
\label{sec:over}

\subsection{Cygnus OB7}
\label{sec:CygnusOB7}

The Cygnus constellation contains several prominent HII regions, OB associations, and massive star clusters. Particularly, the Cygnus OB2 association is notable in the Galaxy because of the large number of massive stars, including Wolf-Rayets. \citet{Reipurth2008} presented an extinction map of the Cygnus molecular cloud complex, providing a complete overview of the star forming regions and young star clusters in Cygnus showing (see their Fig. 1). Their map includes Cygnus-X, the North American Nebula, the Pelican Nebula, and COB7- MC, all of them lying within a distance from the Sun ranging from 600 pc (North American and Pelican Nebula) to 1.7 kpc (Cygnus-X; \cite{Schneider2006}). Considering a distance of 800 pc \citet{Humphreys1978,Dobashi1994,Dobashi1996} estimate a total mass of M $\sim$ 1 $\times$ 10$^5$ M$_{\odot}$.

\begin{figure*}[!ht]
\begin{center}
\includegraphics[width=0.7\textwidth]{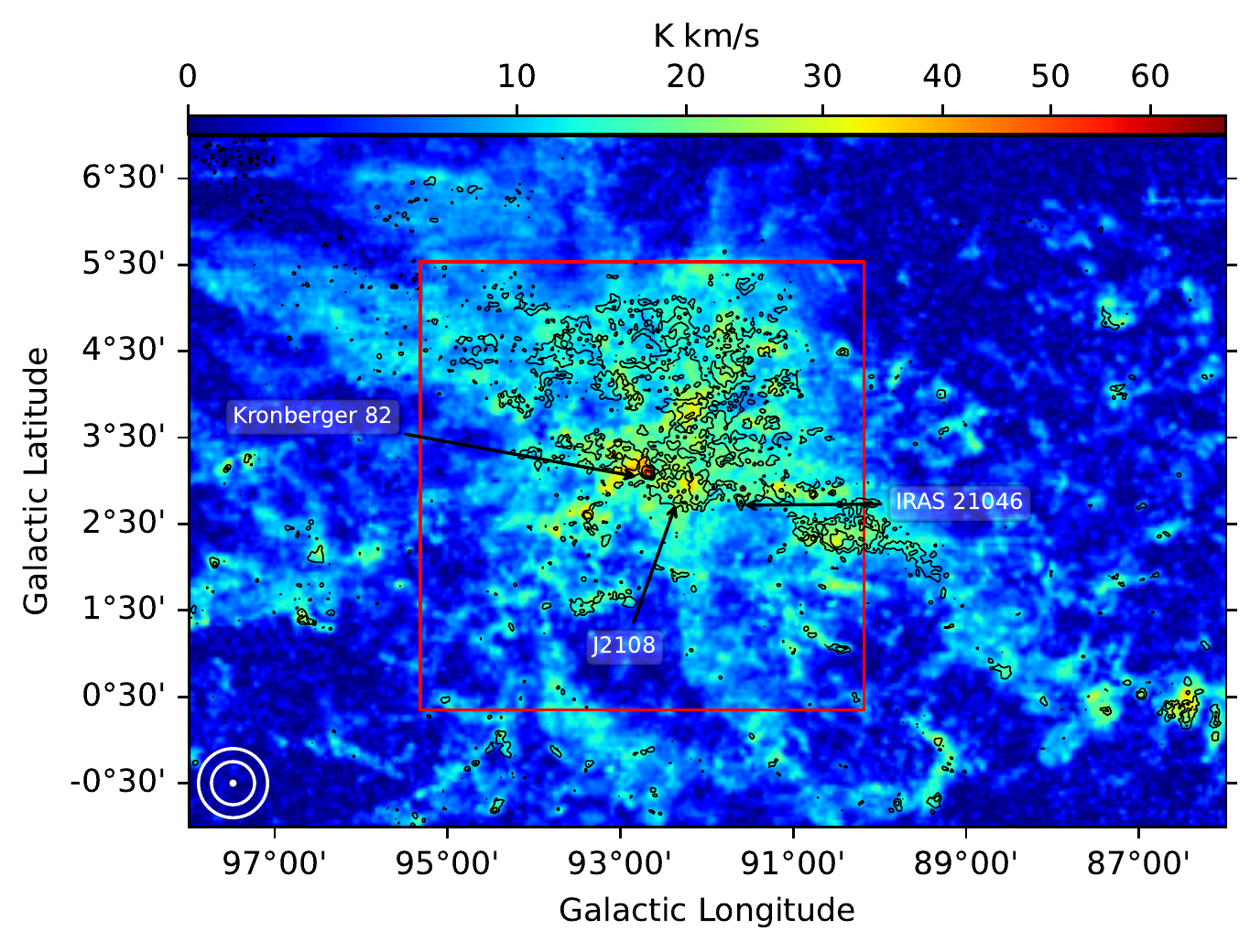}
\end{center}
\caption{$^{\rm 12}$CO(J=2$\rightarrow$1) map for COB7-MC \citep{Nishimura2020} with the $^{13}$CO(J=2$\rightarrow$1) emission overlaid as black contours for different representative velocities ranges (this work). The positions of Kron 82, IRAS 21046, and J2108 are indicated with black arrows. The red square represents the field in Fig. \ref{fig:HI_CO} within which only HI emission was observed. The beams of LHAASO (0.8$^{\circ}$ and 0.5$^{\circ}$) and OPU (2.7') are shown as white circles in the bottom left corner.
}
\label{fig:cygOB7}
\end{figure*}

The large-scale regions of COB7 were surveyed in the infrared by WISE and also partially by GLIMPSE 360 (3.6$\mu$m and 4.5$\mu$m). We show the the WISE RGB image (22$\mu$m + 4.6$\mu$m + 3.4$\mu$m) in Fig.~\ref{fig:CygOB7_intro} (top). Some regions, mainly associated with IRAS sources, are labelled for their easy identification. In the bottom panel of the same figure we also show an image of the $^{12}$CO(J=2$\rightarrow$1) emission obtained with the 1.85 m radio telescope at Osaka Prefecture University (OPU). These observations were presented by \citet{Nishimura2020}, however, they did not perform any analysis of the observations. The morphology of the emission is compatible with the distribution of the emission in the low-resolution $^{13}$CO(J=1$\rightarrow$0) observations (angular resolution of $\sim$ 3') performed with the two 4m-millimetre telescopes of the Nagoya Observatory \citep[][see also Fig. 1 of \citealt{Dobashi2018}]{Dobashi1994}. The $^{12}$CO(J=2$\rightarrow$1) map is the same one investigated and described in this work by adding the corresponding $,^{13}$CO(J=2$\rightarrow$1) observations (see \S~\ref{sec:obs}).

\begin{figure*}[!ht]

\includegraphics[width=0.325\textwidth]{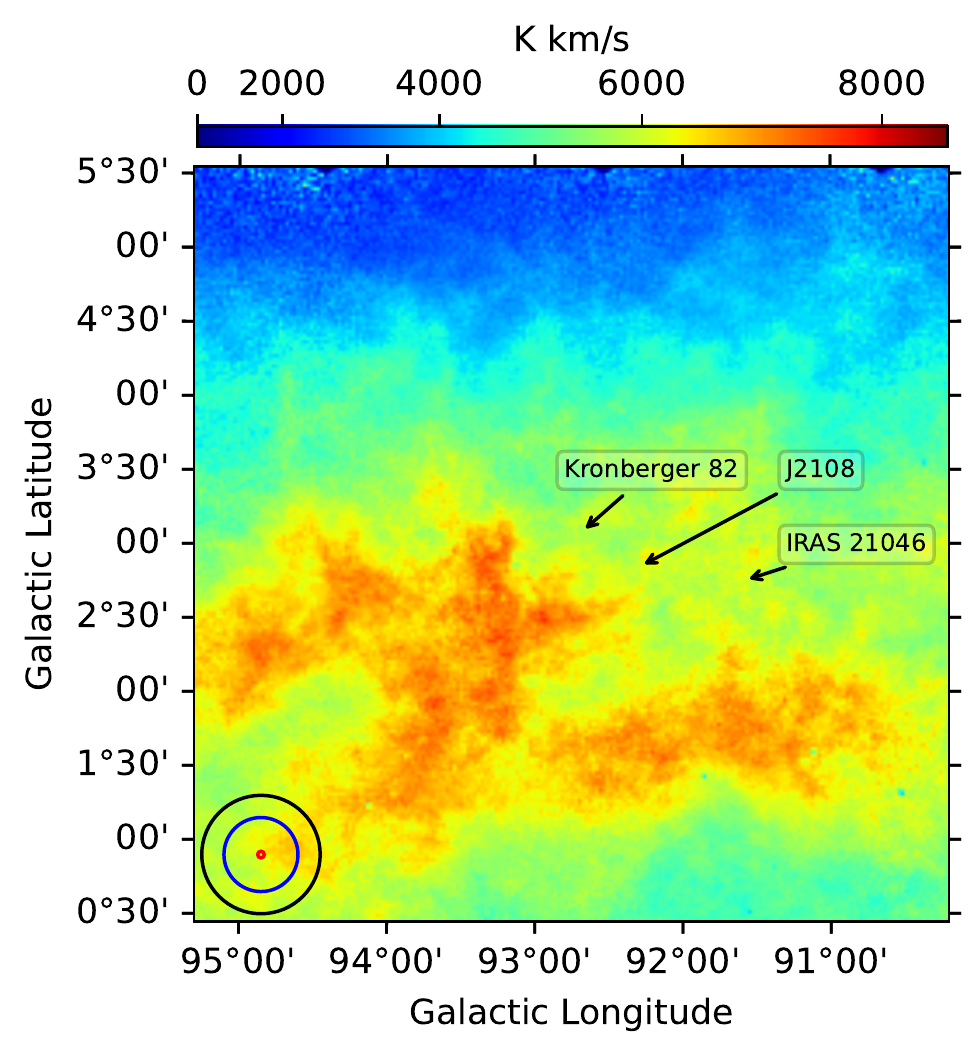}
\includegraphics[width=0.325\textwidth]{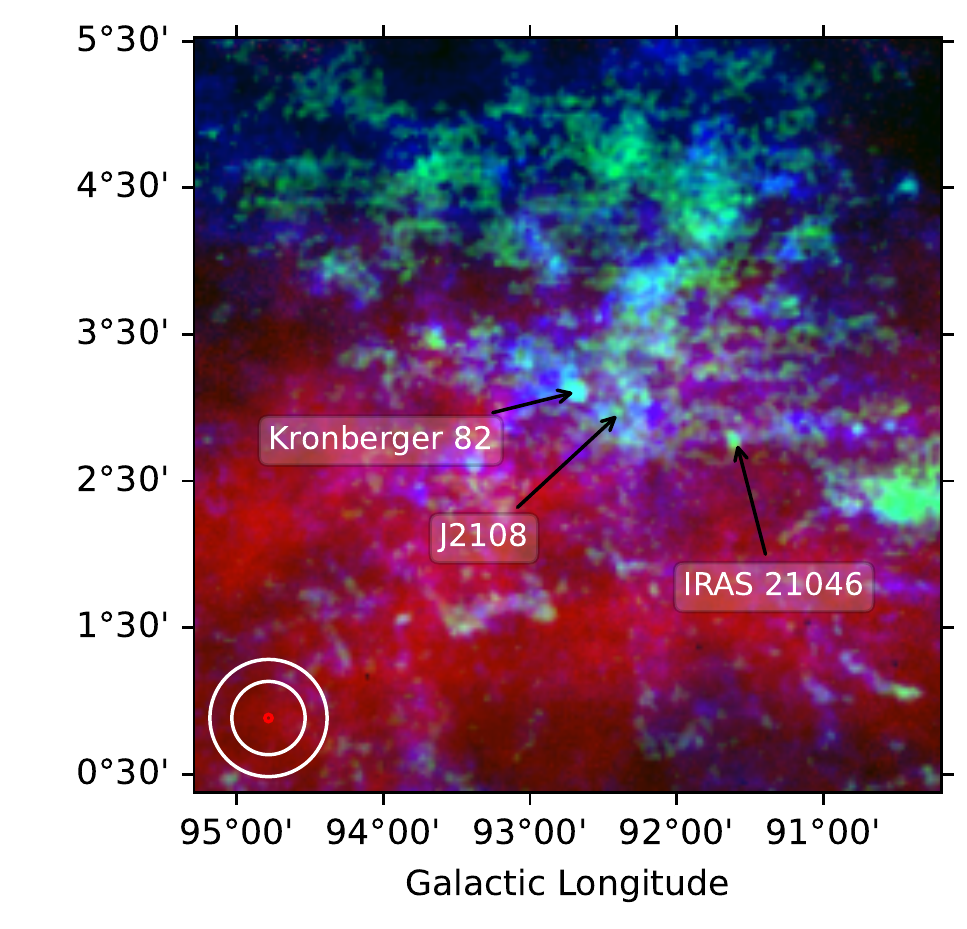}
\includegraphics[width=0.325\textwidth]{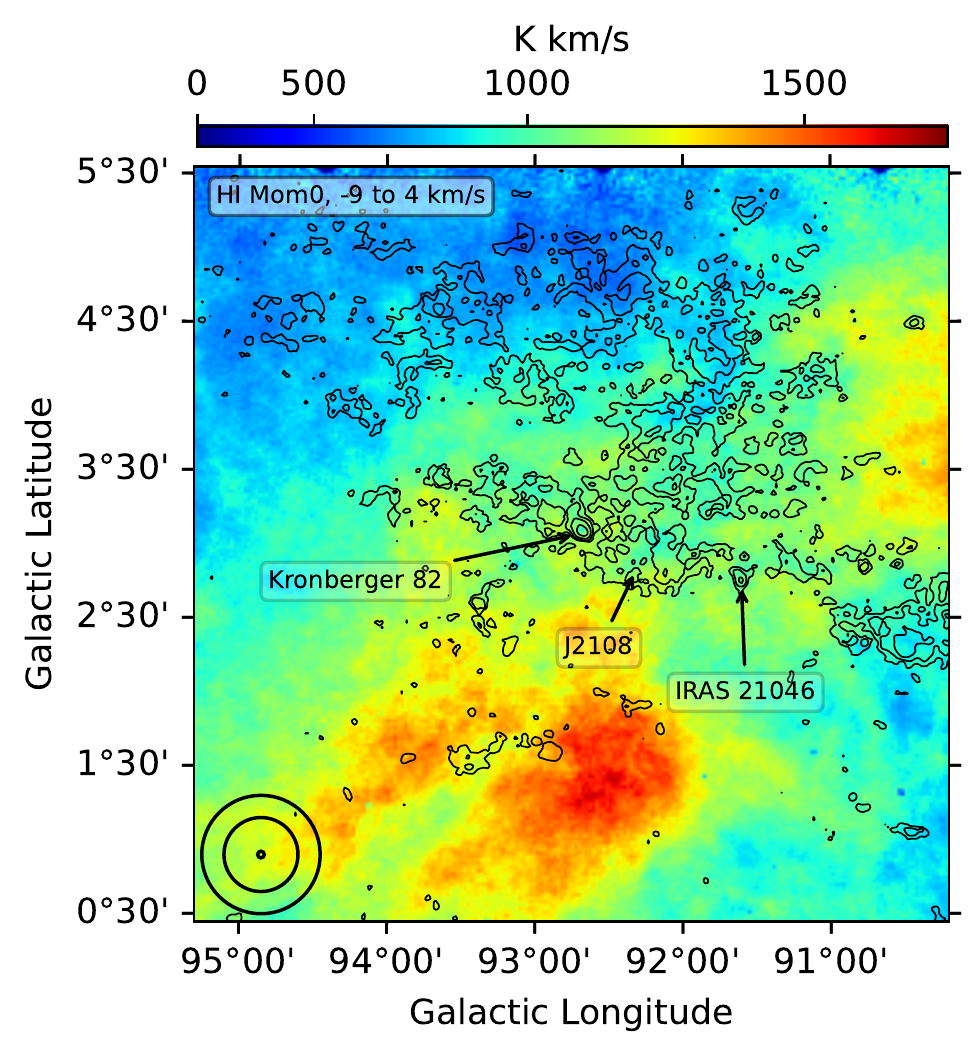}

\caption{\textbf{Left:} DRAO HI (21 cm) line image with J2108 at the center ($\alpha$(2000) = 21$^h$08$^m$36$^s$; $\delta$(2000) = +51$^{\circ}$57'00''). The image was generated considering an integration range between --150 and 50 km s$^{-1}$. The HI image does not cover the whole CO field compared with Fig.~\ref{fig:cygOB7}. \textbf{Center:} RGB (red = DRAO 21 cm, green = $^{13}$CO(J=2$\rightarrow$1), blue = $^{12}$CO(J=2$\rightarrow$1)) image from the same field at the left panel. \textbf{Right:} HI (21 cm) line emission, integrated between --9 and 4 km s$^{-1}$ in the same field of view as the left panel, with $^{13}$CO(J=2$\rightarrow$1) emission overlaid in black contours (integrated between --100 to 80 km s$^{-1}$). The size of images is 5$\times$5 degrees. The y-axis and x-axis are Galactic latitude and longitude, respectively. The beams of LHAASO (0.8$^{\circ}$ and 0.5$^{\circ}$) and OPU (2.7') are shown in circles at the bottom left.}
\label{fig:HI_CO}
\end{figure*}

\begin{figure*}
\begin{minipage}{.5\linewidth}
\centering
\includegraphics[scale=.6]{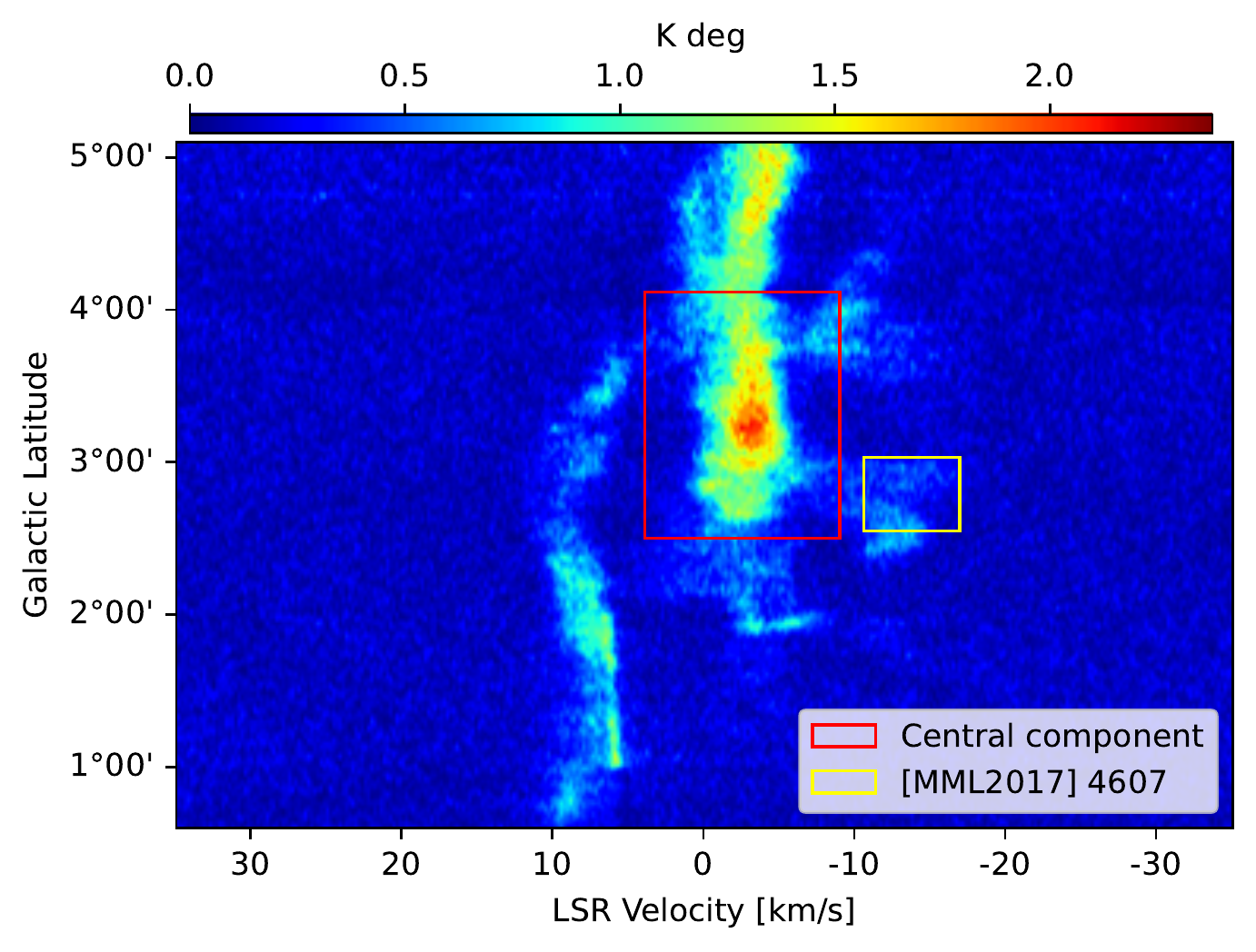}
\end{minipage}%
\begin{minipage}{.5\linewidth}
\centering
\includegraphics[scale=.6]{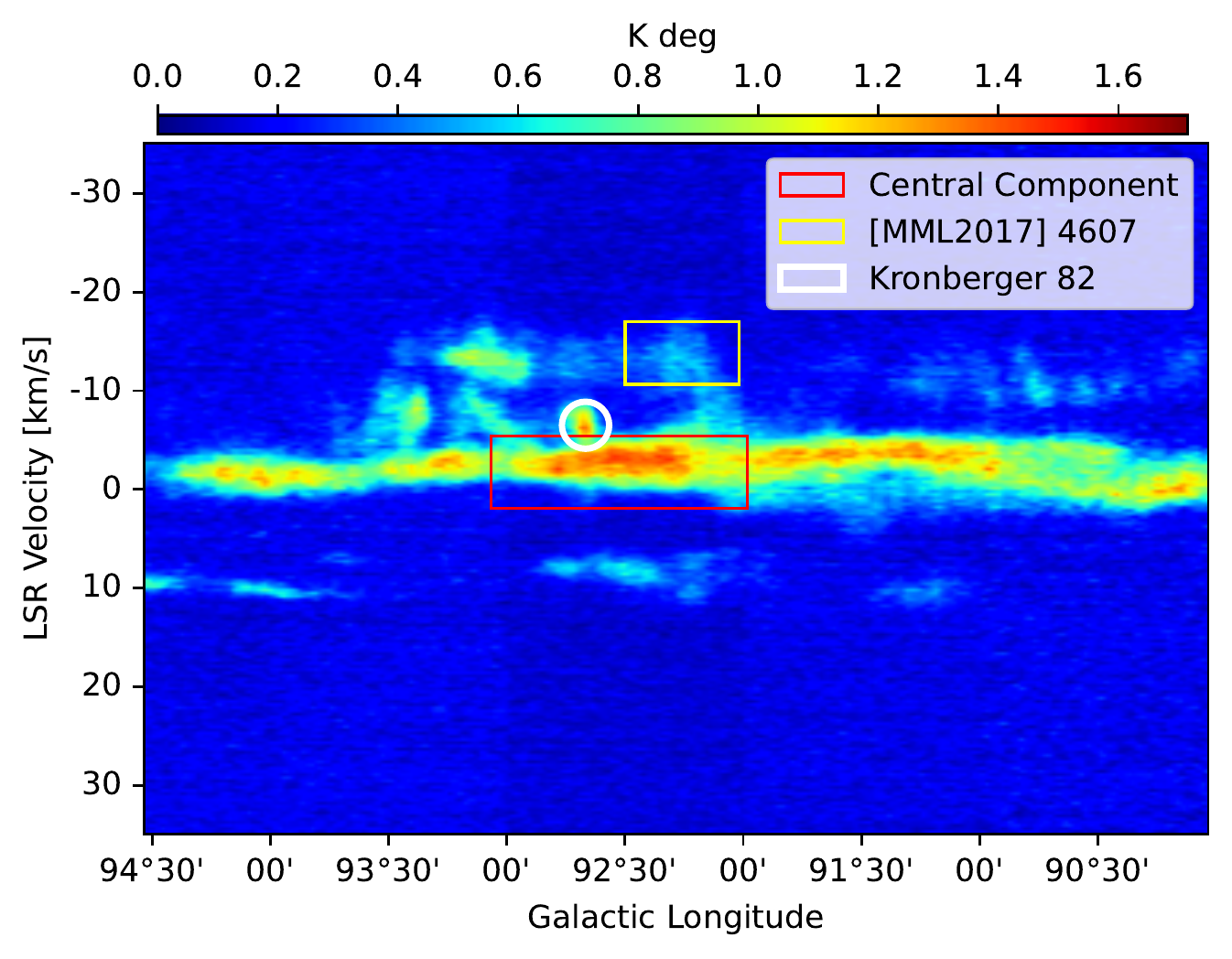}
\end{minipage}\par\medskip
\begin{minipage}{.5\linewidth}
\centering
\includegraphics[scale=.6]{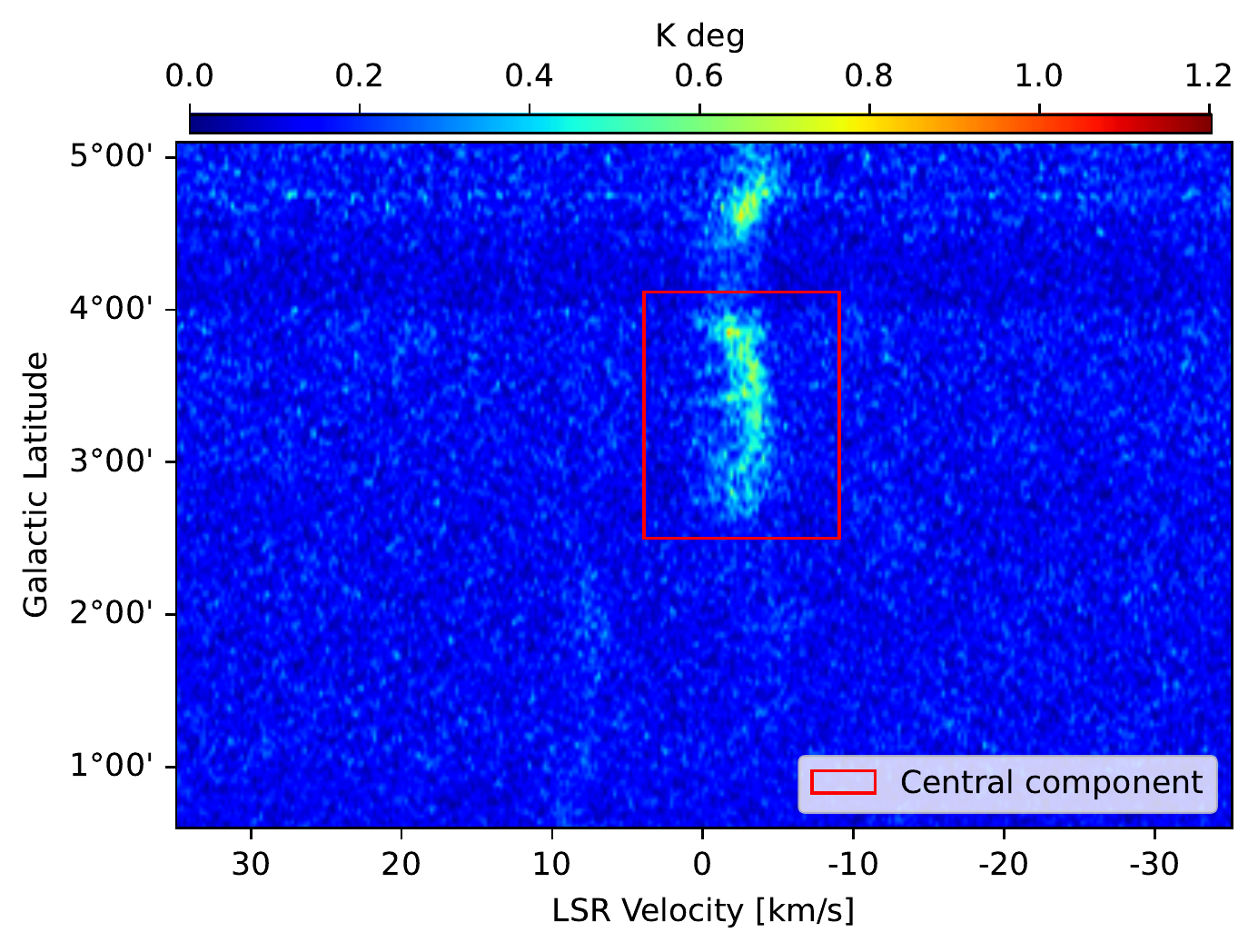}
\end{minipage}%
\begin{minipage}{.5\linewidth}
\centering
\includegraphics[scale=.6]{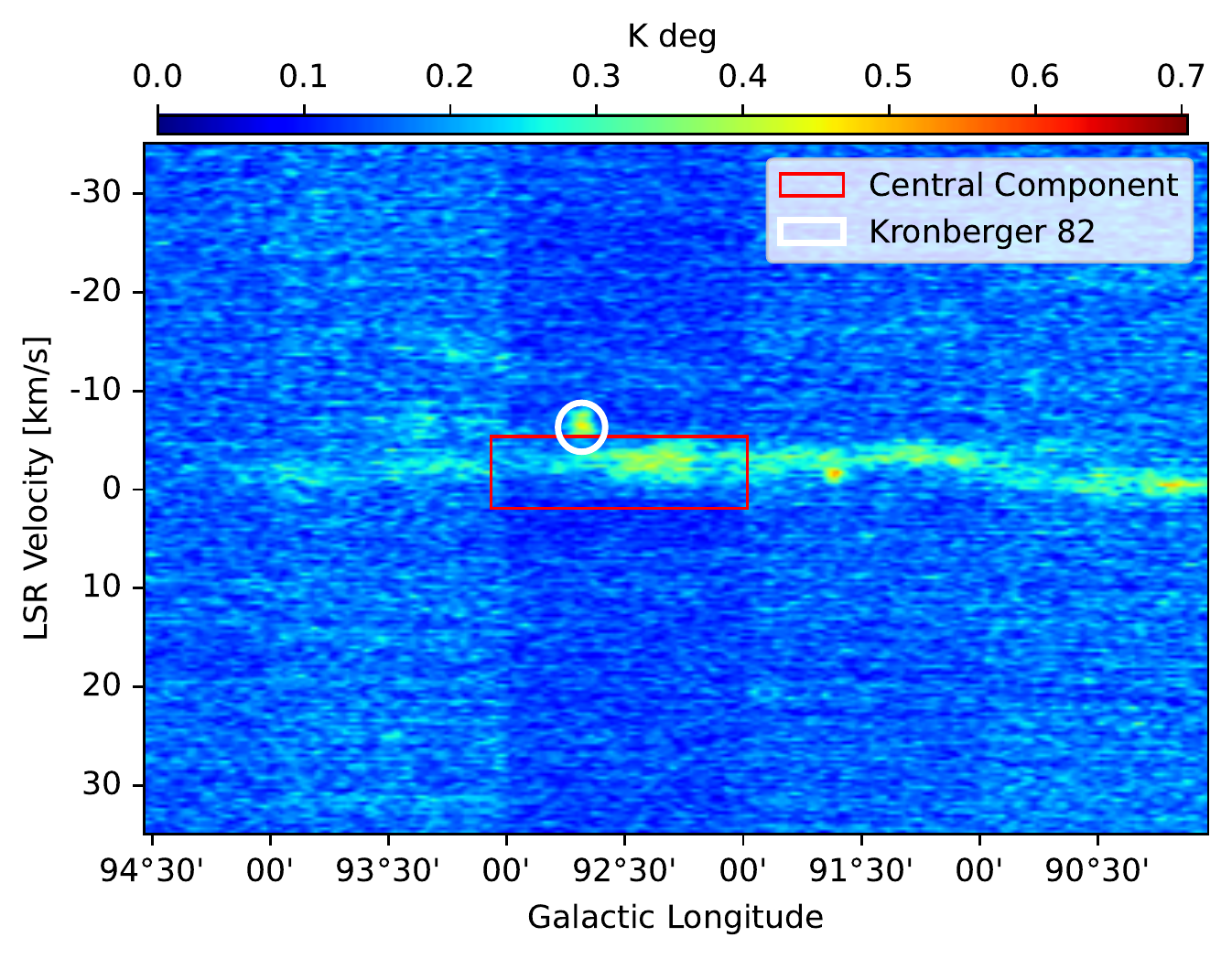}
\end{minipage}\par\medskip
\caption{Galactic longitude and latitude $^{12}$CO(J=2$\rightarrow$1) (top panels) and $^{13}$CO(J=2$\rightarrow$1) (bottom panels) emission maps as function of the V$_{\rm LSR}$ for J2108 vicinity. Data have been integrated into an interval of 0.5$^{\circ}$ around the centre location of LHAASO J2108+5157: $l$ = 92.31, and $b$ = 2.84. The molecular cloud in the red rectangle is [FKT-MC]2022 (V$_{\rm LSR}$ $\sim$ --3 km s$^{-1}$), and the yellow rectangle covers MML (V$_{\rm LSR}$ $\sim$ --13 km s$^{-1}$), which is not observed at $^{13}$CO(J=2$\to$1). The cloud just above the red rectangle in both right panels is Kron 82 (V$_{\rm LSR}$ $\sim$ --7 km s$^{-1}$), which was not considered in the [FKT-MC]2022 size calculation (see \S~\ref{sec:size}).  }
\label{fig:VLSR_pos_CO}
\end{figure*}

\begin{figure*}
\begin{minipage}{.33\linewidth}
\centering
\includegraphics[scale=.55]{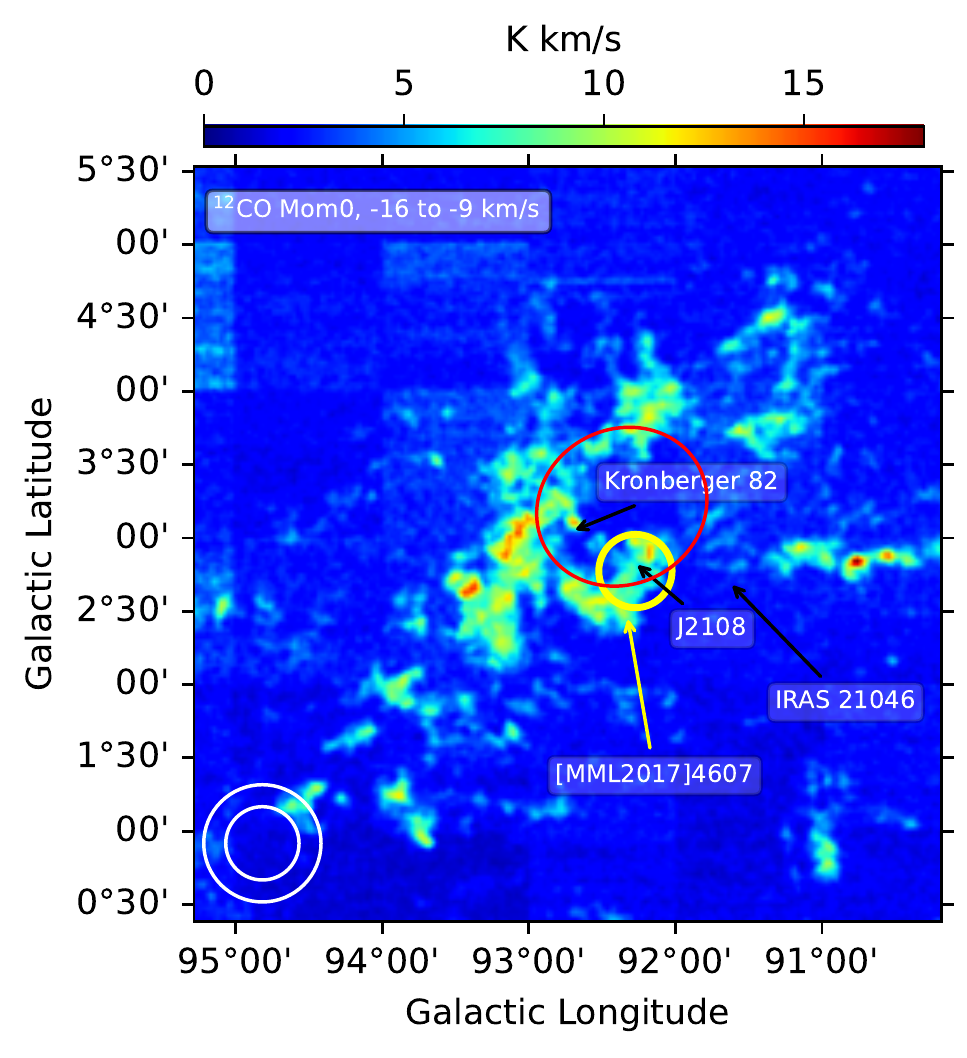}
\end{minipage}%
\begin{minipage}{.33\linewidth}
\centering
\includegraphics[scale=.55]{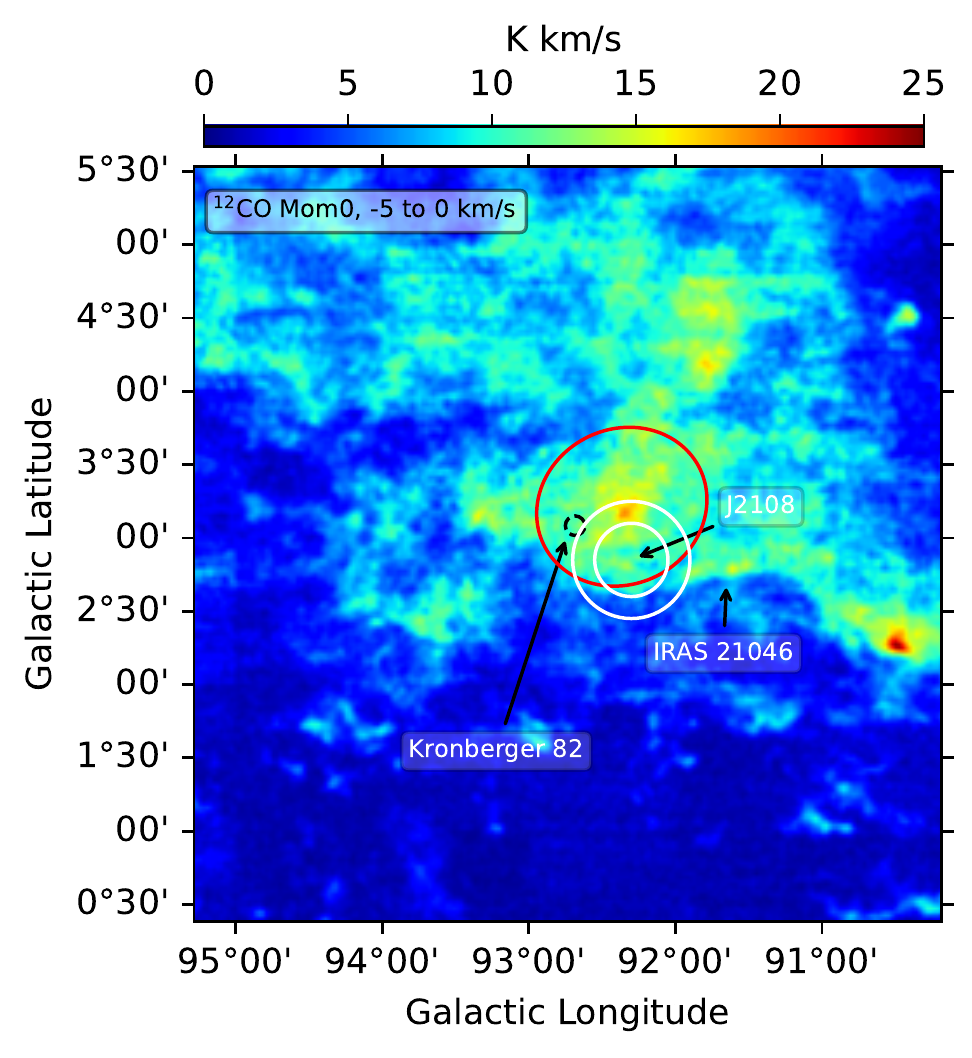}
\end{minipage}%
\begin{minipage}{.33\linewidth}
\centering
\includegraphics[scale=.55]{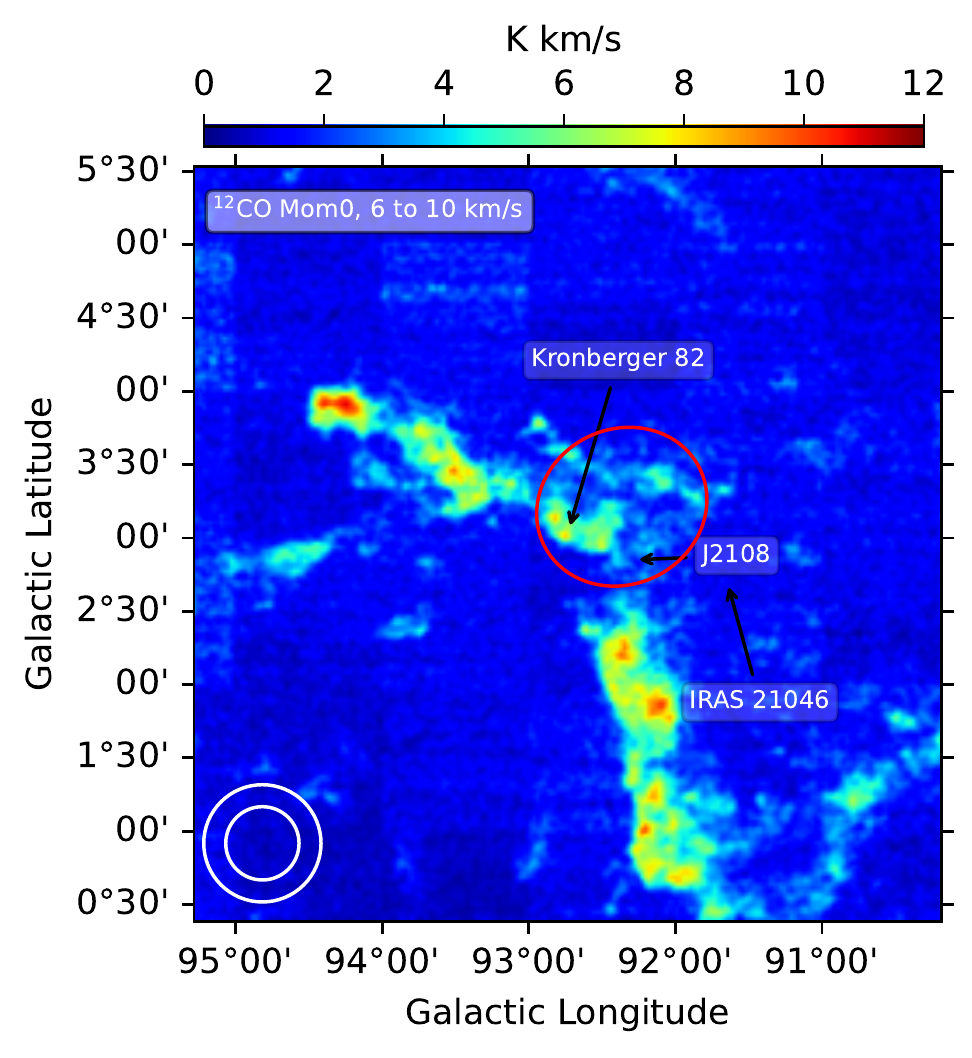}
\end{minipage}\par\medskip
\begin{minipage}{.33\linewidth}
\centering
\includegraphics[scale=.55]{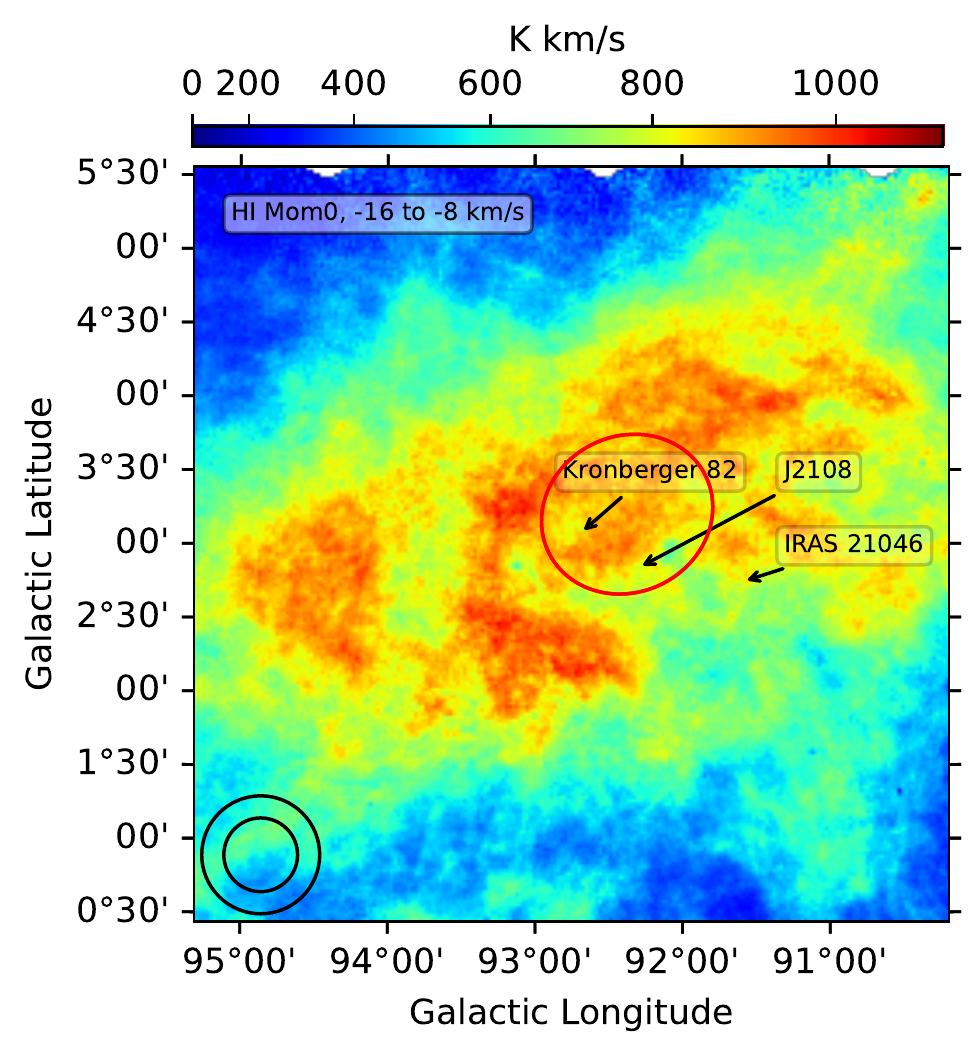}
\end{minipage}%
\begin{minipage}{.33\linewidth}
\centering
\includegraphics[scale=.55]{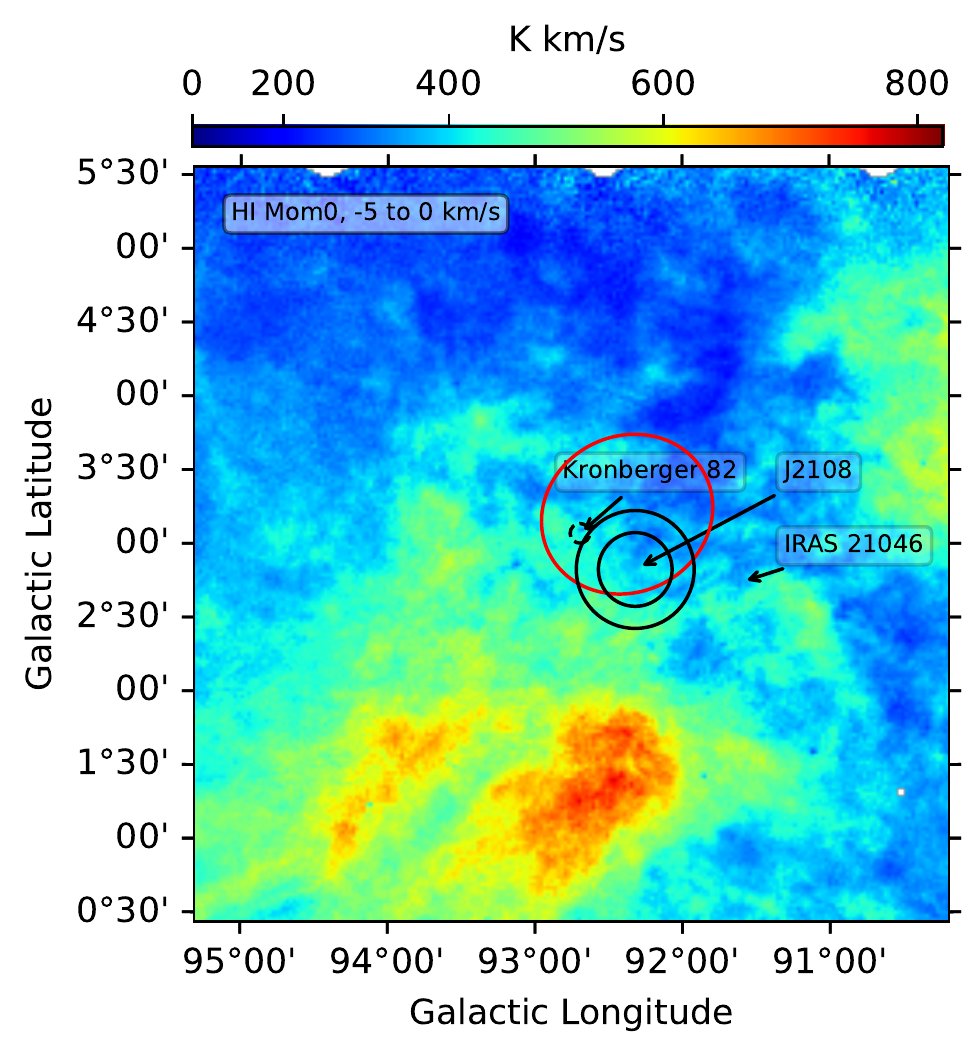}
\end{minipage}%
\begin{minipage}{.33\linewidth}
\centering
\includegraphics[scale=.55]{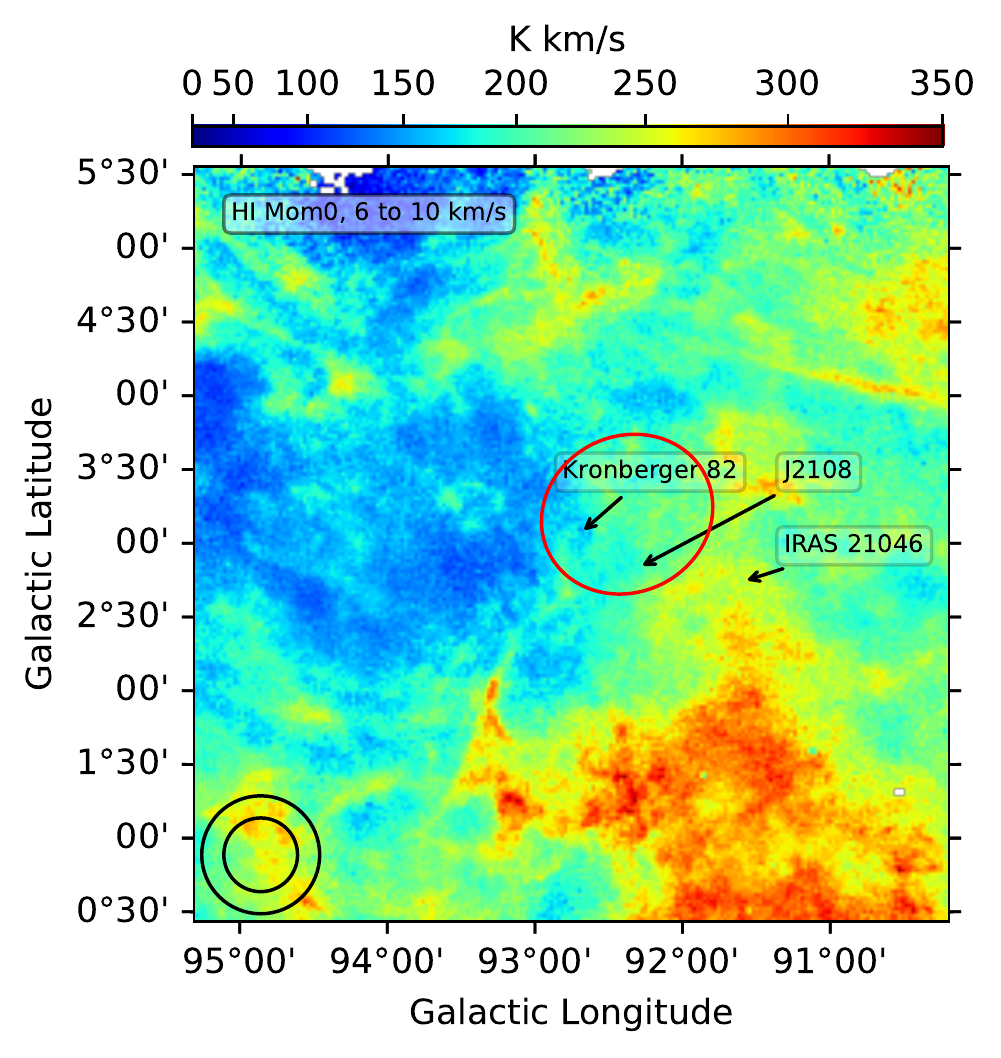}
\end{minipage}\par\medskip
 \caption{Maps of moment 0 for $^{12}$CO (top panels) and HI (bottom panels) emissions. The range of integration in LSR velocity corresponds to the FWHM for the three spectral components of $^{12}$CO emission shown in Fig. \ref{fig:spectra_2deg}. Except for central panels, the beams of LHAASO (0.8$^{\circ}$ and 0.5$^{\circ}$) are shown as circles at the bottom left. For the central panels, these beams are located at centre. The yellow (top-left) and red circles (all panels) correspond to the sizes of MML and [FKT-MC]2022 respectively. Both clouds are inside the resolution of LHAASO. An intersection zone is observed at the center of J2108. In Fig.~\ref{fig:VLSR_pos_CO}, Kron 82 only shows emissions in the galactic longitude maps. Nevertheless, we remove the associated pixels (small black dashed circle at top-centre panel) in the computing of [FKT-MC]2022 size (see \S \ref{sec:size}). Therefore, the FKT-MC]2022 size does not include Kron 82 emission.}
\label{fig:VLSR_pos_HI}
\end{figure*}

\begin{figure}[ht!]
\begin{center}
\includegraphics[width=\columnwidth]{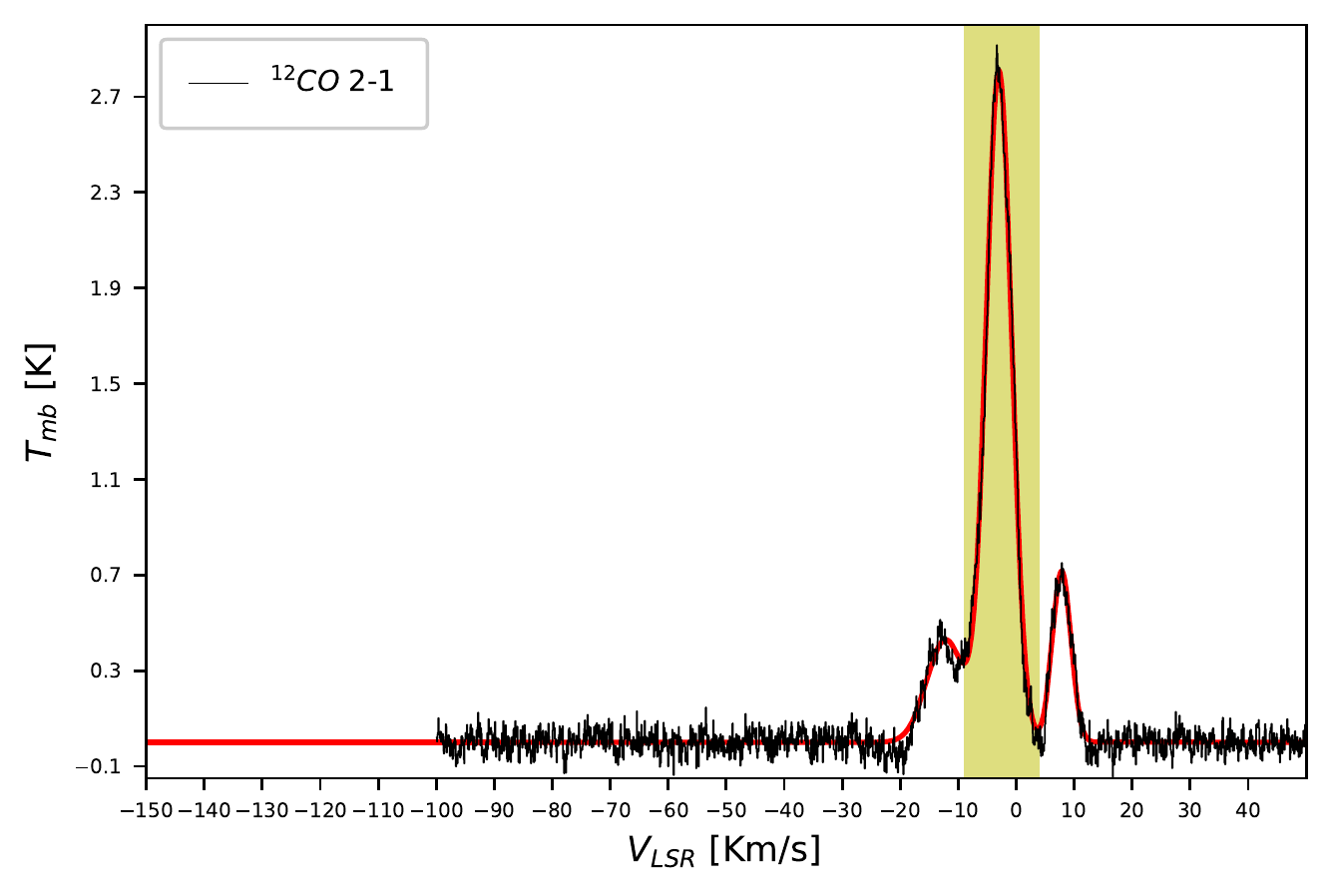}
\includegraphics[width=\columnwidth]{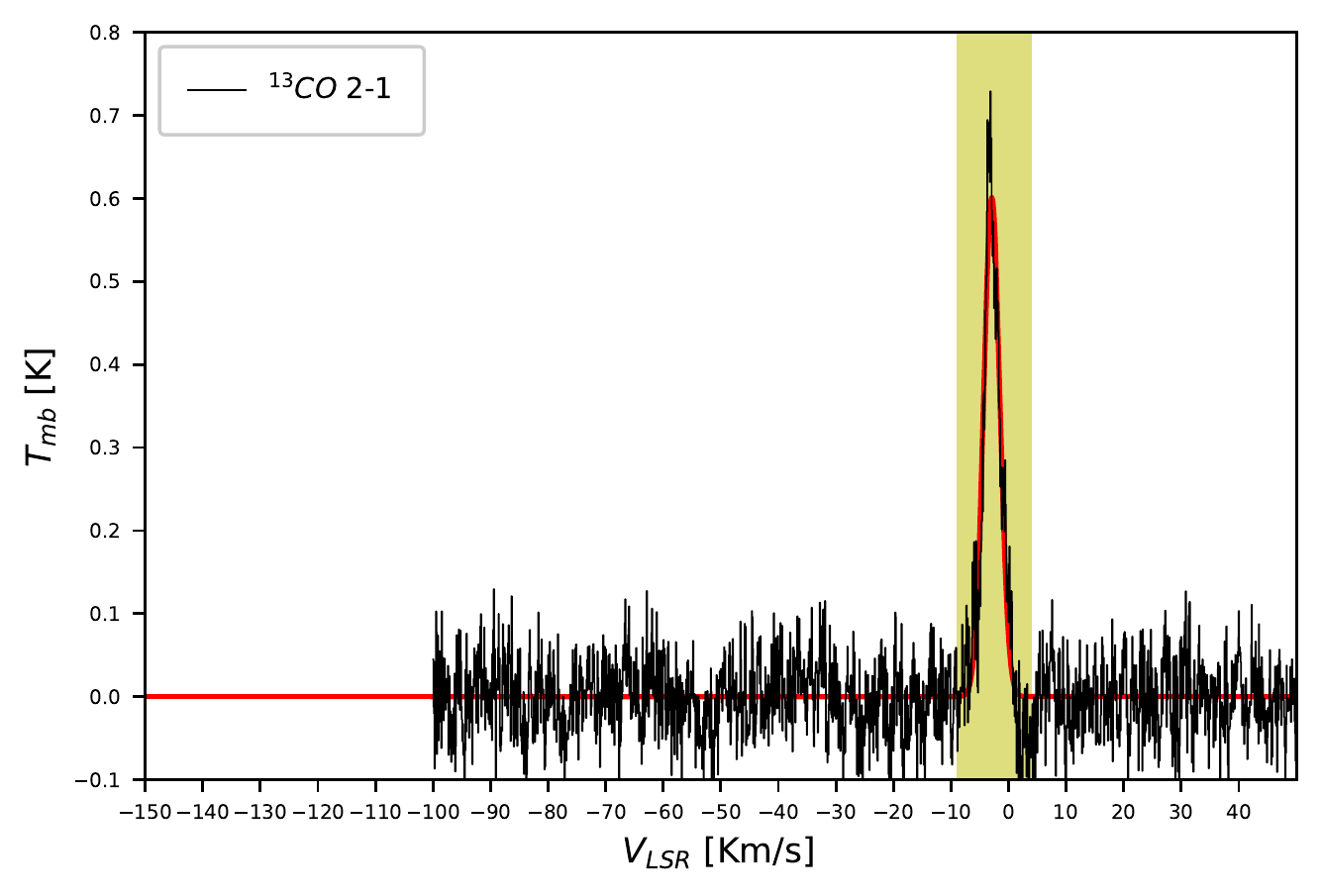}
\includegraphics[width=\columnwidth]{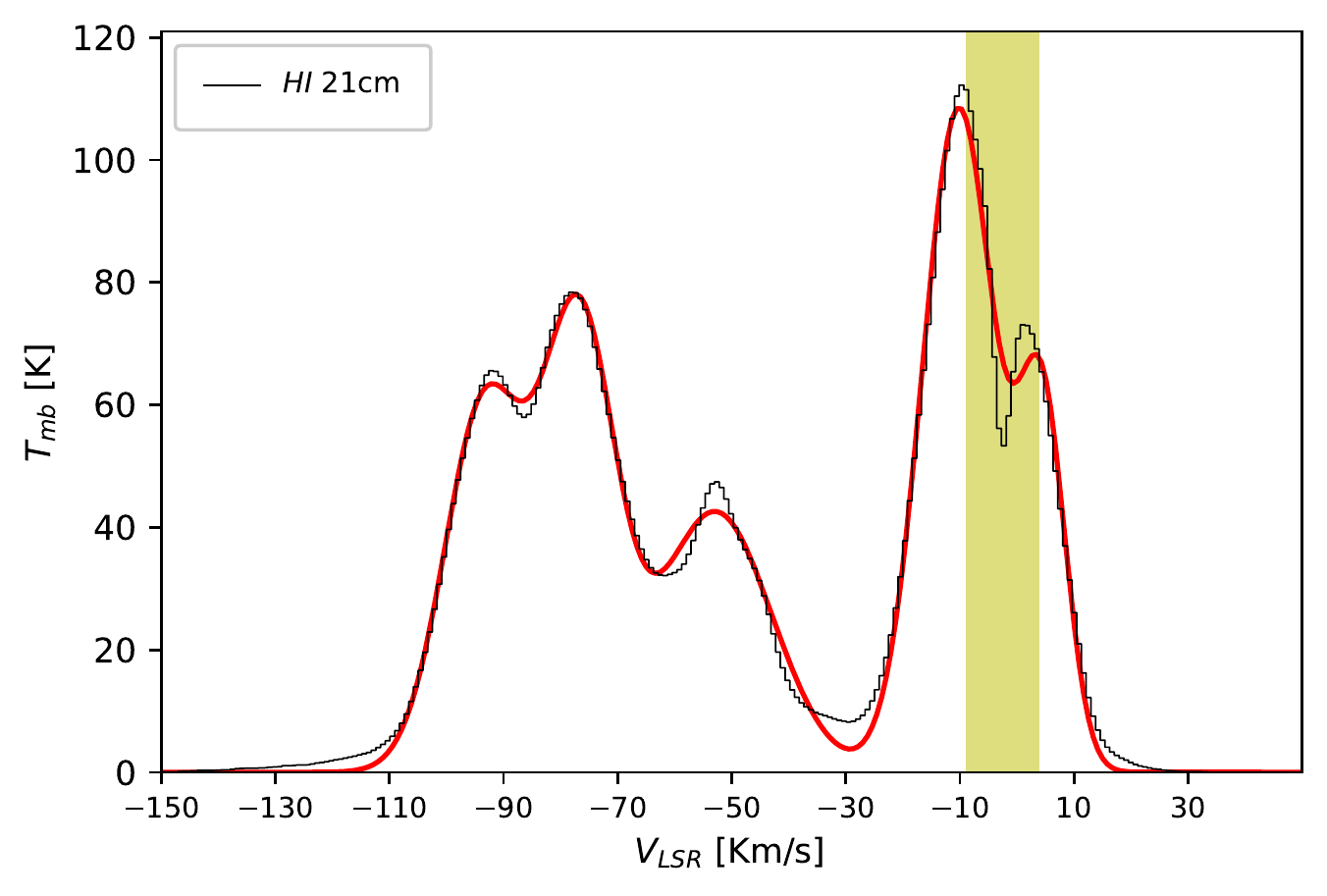}
\end{center}
\caption{Total integrated spectra for the OPU $^{12,13}$CO(J=2$\to$1) and DRAO HI (21 cm) lines observations. The spectra were obtained for the red region depicted in Fig. \ref{fig:VLSR_pos_HI}. The emission map for the different components is shown in Fig.~\ref{fig:VLSR_pos_HI} for CO and HI. The yellow hatched areas fill the velocity range of the main component for $^{12,13}$CO(J=2$\to$1). This range covers part of the main (more intense) for the HI with part of the following right component (see text for details). The components to the left and right of the central line for $^{12}$CO(J=2$\to$1), correspond to COB-7 molecular gas at different LSR velocities. }
\label{fig:spectra_2deg}
\end{figure}

\subsection{LHAASO J2108+5157}
\label{sec:J2108}

J2108 is located at $\alpha$(2000) = 21$^h$08$^m$52.8$^s$; $\delta$(2000) = +51$^{\circ}$57'00'' or l = 92.30$^{\circ}$; b = 2.84$^{\circ}$. The LHAASO observatory detected very high-energy gamma-ray emission in the energy ranges of 25 to 100 TeV and above 100 TeV at 9.5 $\sigma$ and 8.5 $\sigma$ respectively, with an angular resolution of $\sim$0.3$^{\circ}$\citep{Cao2021a}. Therefore, it was postulated to be a PeVatron candidate \citep{Cao2021a,Cao2021b}. The observed morphology is circular with an upper limit on the source size of 0.26 degrees with 95\% confidence level covering a flux at 100 TeV of 0.38$\pm$0.09 CU \footnote{ CU = Crab Units = flux of the crab Nebula at 100 TeV = 6.1 $\times$ 10$^{-17}$ photons TeV$^{-1}$ cm$^{-2}$ s$^{-1}$.}. A power law spectrum with a spectral index of --2.83 best describes the spectral energy distribution. The gamma-ray source has not been identified, no distance has been determined, and there is no known TeV counterpart. To date, neither HAWC (sensitive to a few hundred GeV to above 100 TeV; \citealt{Albert2020b}) nor the Tibet-AS$\gamma$ Observatory (sensitive to energy above 100 TeV; \citealt{Albert2020b}) has reported emission at that energies.

The NASA-FERMI source 4FGL J2108.0+5155 at J(2000) RA = 317.02$^{\circ}$ and Dec = 51.92$^{\circ}$) has a separation of 0.13$^{\circ}$ from the sub-PeV emission observed by LHAASO being the only gamma source in the vicinity\citep{Abdollahi2020}. A symmetric Gaussian describes the morphology of the GeV emission with a width of 0.48$^{\circ}$ \citep{Cao2021b}. Assuming that the spectral energy distribution of the source extends to the VHE energies, extrapolating the spectrum to these energies leads to a flux lower by a factor of 10 than the one observed by LHAASO\citep{Abdollahi2020, Cao2021b}. The nearest X-ray source, an eclipsing binary RX J2107.3+5202 \citep{Motch1997}, is separated by 0.3$^{\circ}$. 

According to LHAASO, the gamma-ray emission from J2108 can be described by leptonic and hadronic origins, although there are no known SNRs and PWNe within 0.8$^{\circ}$ of the centre of the sub-PeV emission. Even though there is no known pulsar in this region to support the leptonic scenario, the possible contribution of a yet unknown pulsar cannot be ignored. On the other hand, \citet{Cao2021b} suggest an association with the molecular cloud [MML2017]4607 (MML hereafter) with radius = 0.24$^{\circ}$, mass = 8469 M$_{\odot}$, and distance $\sim$ 3.3 kpc \citep[3.28 kpc;][]{Miville2017}, favouring the hadronic emission. This result is based on the $^{12}$CO(J=1$\rightarrow$0) survey of \citet{Dame2001}, where the optical depth ($\tau$) is optically thick and does not include emission from inner regions. In any case, although there is no consensus on the distance, it is worthwhile to make the same suggestion but considering optically thin observations, high-density tracers of clumps, and other distances to COB7-MC such as those reported by \citet{Humphreys1978,Schneider2006} from 800 pc to 1.7 kpc instead of 3.3 kpc.

A lepto-hadronic model in sub-PeV emission environments associated with giant molecular clouds without a strong pulsar or supernova remnant as in J2108 is described by \citet{Kar2022}. This model assumes collisionally accelerated electrons and protons injected into the local interstellar medium from past supernova explosions thousands of years ago. This model can be tested by combining high-resolution HI (unavailable) with X-rays (leptonic emission), shocked HII gas tracers such as H$\alpha$, forbidden lines (optical supernova remnants), and radio-continuum (RC) observations at 20 cm (synchrotron and pulsars) and 3.6 cm (Bremsstrahlung). Successful examples of the discovery, confirmation, and study of supernova remnants interacting with molecular clouds include SN 1006, RX J1713.7-3946, 3C400.2, G350.0-2.0, IC443 G352.7-0.1, and others \citep[e.g][and references therein]{Sano2022,Fukui2021,Fukui2012,Erging2017,Karpova2016,ToledoRoy2014,Jiang2010,Schneiter2008,Albert2007,Ambrocio2006}. Detecting a supernova (and its remnant) in J2108 is crucial for understanding the contribution of leptonic (and possibly hadronic) emission to the observed sub-PeV emission.      

\section{Observations}
\label{sec:obs}

We report observations of COB7-MC carried out with the 1.85 m radio-telescope \citep{Nishimura2020,Onishi2013} of the OPU at Nobeyama Radio Observatory. The observations were conducted from February to May 2011. A receiver for observation at 230 GHz was used to produce $^{12}$CO(J=2$\rightarrow$1) and $^{13}$CO(J=2$\rightarrow$1) maps with an angular resolution of $\sim$3$\arcmin$ covering a spectral window from -100 to 80 km s$^{-1}$ in the local standard of rest (LSR). The data were calibrated with the standard procedure used for previous data sets obtained with the same telescope \citep{Onishi2013,Nishimura2015}. The RMS value of the noise is $\sim$0.3\,K at a velocity resolution of 0.3\,km\,s$^{-1}$.

We retrieve atomic hydrogen (HI) 21 cm line observations from the Dominion Radio Astrophysical Observatory (DRAO\footnote{DRAO is part of the Canadian Galactic Plane Survey Project (CGPS). \url{https://www.cadc-ccda.hia-iha.nrc-cnrc.gc.ca/en/search/#resultTableTab}}) archives \citep{Taylor2003}, at a Galactic longitude and latitude that includes the position of J2108. These observations were taken on August 20, 2001, with a resolution of 1 arcminute at 1420 MHz. The data set is projected into a 1024$\times$ 1024, 18 arcsec pixel mosaic image, with 0.82 km s$^{-1}$ velocity channels. The RMS noise of the brightness temperature for an empty channel is between 2.1 and 3.2 K.


\section{Methodology: The density of nucleons; n(H)}
\label{sec:methodology}

Assuming that no external astronomical source such as the PeVatron is found in the region where the sub-PeV emission is observed, a molecular cloud or clump in a molecular environment, where the neutral pion decay process may occur, is a good candidate to produce this (hadronic) emission. In this scenario, the hadronic contribution (N$^{\rm had}_{\gamma}$) may be calculated from the observed total flux of gamma-ray emission.

\begin{equation}
\rm{N^{\rm obs}_{\gamma} = \rm{N^{\rm had}_{\gamma}} + N^{\rm lep}_{\gamma}} ,
\label{eq:eqnmodelgeneral}
\end{equation}

as

\begin{equation}
\label{eq:eqnmodel}
\rm{N^{had}_\gamma} = \rm{N^{obs}_\gamma} - \rm{N^{lep}_\gamma} \propto c~ n(\rm{H}) N_{\rm P}(\rm CR),
\end{equation}

with:

\begin{equation}
\rm {N^{\rm lep}_{\gamma}} \propto \rm{N^{\rm obs}_{X-Ray}\Bigg[\frac{N(CMB)}{B^2}\Bigg]}.
\label{eq:eqnmodel_lep}
\end{equation}

In these equations, c is the speed of light in vacuum, N$_{\gamma}$ is a differential flux $\frac{dN}{dE}$ corresponding to total emission  (hadronic and leptonic) expressed in units of $\rm TeV^{-1} \ cm^{-2} \ s^{-1}$, n(H) is the numeric density of the nucleons (protons) involved in the hadronic emission (N$^{\rm had}_{\gamma}$), N$_{\rm P}$(CR) corresponds to cosmic ray counts (CR) in hadronic emission \citep[e.g][]{Amenomori2021b,Ackermann2011}, N$^{\rm obs}_{\rm X-ray}$ is the observed X--ray count, N(CMB) is the photon density from the cosmic microwave background, and B is the ISM magnetic field. The X--ray emission is required to calculate the leptonic contribution using the inverse Compton effect. The Eq.~\ref{eq:eqnmodel} is supported by standard models \citep{Kelner2006} and has been used effectively to SNR RX J1713.7-3946 \citep{Fukui2021,Fukui2012}. Observations of leptonic emission can constrain the value of N$^{\rm lep}_{\gamma}$ \citep[e.g. X--rays;][and references therein]{Fukui2021}, and n(H) can be calculated from the total proton column density, N(H), obtained from CO and HI observations as

\begin{equation}
\label{eq:eqnNH}
\rm N(H) = 2 N(H_2) + N(HI),
\end{equation}

where N(H$_2$) is calculated from the CO column density, N(CO). To obtain a complete understanding of the protons in the ISM, the column density of HI, N(HI), must be included in this equation\citep[see][and references therein]{Fukui2021, Fukui2012, Ade2011}. Then, in the standard way (see Appendix~\ref{appendix:ap1}) 

\begin{equation}
\label{eq:eqnNH2}
 \rm N(H_2) = 5 \times 10^{5} \ N \left( ^{13}CO \right),
\end{equation}

\begin{equation}
\label{eq:eqnNHI}
    \left[ \frac{N({\rm HI})}{\rm cm^{-2}} \right]= 1.823 \times 10^{18} \left[ \frac{\int T_{\rm MB}^{HI} dv}{\rm K \ km \ s^{-1}}\right],
\end{equation}

\noindent where N($^{13}$CO) is the $^{13}$CO column density (derived from J=2$\rightarrow$1 line), and $T_{\rm MB }^{ HI }$ is the brightness temperature averaged over the main beam as a result of a Gaussian fit of the HI emission spectrum. Using the relationship between column and number density and assuming a constant number density of the cloud, we can estimate the number density as follows:

\begin{equation}
\label{eq:eqnnH}
{\rm n(H)} = l^{-1} \rm{N(H)},
\end{equation}

\noindent where $l$ is the physical size of the region under consideration.

The latter requires a careful search for the molecular clump and the determination of a suitable size, mass and distance, because the calculation of these parameters is fundamental for determining the density of the nucleons. This computing is particularly hard for COB7 due to its complex morphology, since the size, morphology (mainly), and distance are significant sources of uncertainty.

\section{Results and Discussion}
\label{sec:results_discussion}

In Fig.~\ref{fig:lhaaso_opt} (top-left) we present close-up of the $^{12}$CO(J=2$\to$1) images from Fig.~\ref{fig:CygOB7_intro}, but centred on J2108. The circles cover regions of interest (ROIs) with diameters of 0.7, 1.5, and 2.0 degrees, where all source candidates for PeVatron \citep{Cao2021b} including IRAS 21046+5110 are located. Kron 82 and IRAS 21046 are marked 1 and 2, respectively, and are the closest star clusters to J2108. The WISE RGB image (red = 22$\mu$m, green = 4.6$\mu$m blue = 3.4$\mu$m) is shown as a one-degree square ($\sim$ four times the upper limit of LHAASO KM2A PSF for J2108 \citet{Cao2021b} inset in the upper right panel. The Kron 82 emission stands out, and IRAS 21046 does not appear in the field. In the images from WISE, IRAS 21046 is fainter than Kron 82, its emission is not as spectacular, and it is further away from J2108 than Kron 82 (at a distance of $\sim$ 0.7$^{\circ}$ versus $\sim$ 0.4$^{\circ}$; a reason to be discarded by \citet{Cao2021b}. 

There are no precise molecular emissions and counterparts in the centre of J2108, particularly in the upper limits emission extension circles at the upper edge at GeV (NASA-FERMI) and sub-PeV (LHAASO) emissions \citep[see Fig. 4 of][which includes observations from \Citealt{Dame2001}]{Cao2021b}. Kron 82 shows a 20 cm radio continuum (RC; \cite{Taylor2003}) in agreement with the 20 cm RC emission from NVSS \citep{Condon1998}, but no CO emission is observed.

In contrast, Kron 82 is a bright source on the map in our OPU $^{12,13}$CO(J=2$\to$1) emission. Kron 82 is the closest optical/radio counterpart to J2108. If the Cygnus OB2 star cluster is the PeVatron in the Cygnus cocoon, it is worth investigating and confirming whether Kron 82 could be the counterpart of J2108. Therefore, we reinforce the idea that a molecular clump or Kron 82 is the gamma-ray emission candidate. Nevertheless, there is no extended X-ray evidence in the literature for stellar winds or similar processes in massive stars at several degrees. 

In Fig.~\ref{fig:lhaaso_opt} (bottom left) we show a zoom of Kron 82 in an RGB image (NVSS in red, OPU $^{12}$CO(J=2$\to$1) in green, and Spitzer-IRAC at 4.5$\mu$ in blue). As an inset (bottom right), another RGB image (NVSS in red, WISE 22$\mu$m in green, and Spitzer--IRAC 4.5$\mu$m in blue) is shown. Kron 82 is the only object with NVSS emission in the vicinity of J2108 at least in several degrees. 

We present the OPU $^{12,13}$CO(J=2$\to$1) study and gamma-ray modelling for Kron 82 in Appendix ~\ref{appendix:ap3}. This appendix shows that although Kron 82 requires a cosmic-ray energy (proton) of $\sim 10^{45}$ erg to produce the reported LHAASO emission, it is located outside the radial angular extension of J2108. Thus, discarding Kronberger 80 (distance $\sim$ 5 kpc; \citealt{Cao2021b}), IRAS 21046 (distance $\sim$ 600 pc; \citealt{Kumar2006}) and Kron 82 (due to their stellar content, mass and angular separation from J2108; distance $\sim$ 1.6 kpc), a molecular cloud near J2108 could be the best option to produce the observed sub-PeV emission. However, this molecular cloud could not be MML (distance of $\sim$ 3.3 kpc), as stated in \S~\ref{sec:J2108}.

\subsection{Cygnus OB7 and the vicinity of LHAASO J2108+5157}
\label{sec:Cyg0B7}

\subsubsection{HI and CO emission}
\label{sec:vicinity}

In Fig.~\ref{fig:cygOB7}, we show the integrated $^{12}$CO(J=2$\to$1) emission (in colour) in the velocity range between -100 and 80 km s$^{-1}$) with the integrated $^{13}$CO(J=2$\to$1) emission (same velocity range) superimposed as black contours. The $^{13}$CO(J=2$\to$1) emission predominates in the north, especially in the NE, and is almost absent in the south, where the $^{12}$CO(J=2$\to$1) emission remains. The red square marks the region shown in Fig.~\ref{fig:HI_CO}, the only one with HI reported. In this Fig.~we show the DRAO HI 21 cm map (left), an RGB image (center) with DRAO 21 cm in red, $^{13}$CO(J=2$\rightarrow$1) in green, $^{12}$CO(J=2$\rightarrow$1) in blue, and a HI 21 cm map integrated between --9 and 4 km s$^{-1}$ (corresponding to the brightest spectral component of the region; right), overlaid with the $^{13}$CO(J=2$\to$1) emission. The positions of J2108 (LHAASO) and Kron 82 are labelled. The extent of HI and CO gas is huge, but neither J2108 nor Kron82 have a clear clump or counterpart in HI. Remarkably, we observed a strong anti-correlation between the $^{13}$CO(J=2$\to$1) and the HI maps. Furthermore, the $^{12}$CO(J=2$\to$1) emission is more extended than the $^{13}$CO(J=2$\to$1).

To identify molecular clouds near J2108, we show $^{12,13}$CO(J=2$\to$1) maps of Galactic latitude and longitude as a function of V$_{\rm LSR}$ in Fig.~\ref{fig:VLSR_pos_CO}. We detect MML at V$_{\rm LSR}$ $\sim$ --13 km s$^{-1}$ (yellow rectangle) and a gas parcel with V$_{\rm LSR}$ $\sim$ --3 km s$^{-1}$, called central component and enclosed with a red rectangle (see \S~\ref{sec:size}). We observe this cloud ([FKT-MC]2022 below) at $^{\rm 13}$CO(J=2$\to$1) (optically thin emission), unlike MML. Therefore, [FKT-MC]2022 and MML are molecular clouds near J2108, but [FKT-MC]2022 is detected by its optically thin emission.

We can see in Fig.~\ref{fig:VLSR_pos_CO} the $^{\rm 12}$CO(J=2$\to$1) emission associated with Kron 82 at V$_{\rm LSR} \sim -6$ km s$^{-1}$ (white circle). To exclude this contribution, we integrate between -5 and 0 km s$^{-1}$ and present a $^{12}$CO(J=2$\to$1) moment-0 map in the top central panel of Fig~\ref{fig:VLSR_pos_HI}. We identify [FKT-MC]2022 in the central region of the map by fitting a 2-dimensional Gaussian. We estimate a central position of $l =$ 92.4$^{\circ}$ and $b =$ 3.2$^{\circ}$ with corresponding standard deviations $\sigma_l$ = 0.50$^{\circ}$ and $\sigma_b$ = 0.45$^{\circ}$ with a P.A. = 19 deg. Using these fitted values, we estimate the angular extent of the cloud at FWHM from the central position, shown as a red ellipse in Fig~\ref{fig:VLSR_pos_HI}. Finally, if we consider a beam of 2.7$\arcmin$ $\times$ 2.7 $\arcmin$, we obtain a deconvolved size of 1.1 $\pm$ 0.2$^{\circ}$ \citep[][]{Nishimura2020}.

In Fig.~\ref{fig:VLSR_pos_HI} we show moment 0 maps for $^{12}$CO(J=2$\to$1) and HI emissions for three different ranges of LSR velocity: --16 to --9, --5 to 0 and 6 to 10 km s$^{-1}$. The yellow circle in the top-left panel shows the vicinity of J2108 at 0.5$^{\circ}$ where MML is marked. The red ellipse covers [FKT-MC]2022 to its full extent (size of 1.1 $^{\circ}$) in all panels. Except in the central panels, we show the two ``beam sizes" of LHAASO at 0.8$^{\circ}$ and 0.5$^{\circ}$ the bottom left, respectively. In central panels, these ``beams" appear centred in J2108. Comparing the position of [FKT-MC]2022 with the LHAASO beams and MML, [FKT-MC]2022 can also be considered a candidate for producing the sub-PeV emission.

We show in Fig.~\ref{fig:spectra_2deg} the average spectra of the $^{12,13}$CO and HI line emissions obtained from the angular extension of [FKT-MC]2022. The $^{12}$CO spectrum shows three main velocity components, denoted as left, centre (the brightest), and right (see Tab.~\ref{table:adj1}). We have performed a Gaussian fit for the three components. At V$_{\rm LSR} \approx$ --2.9 km s$^{-1}$ to --3.0 km s$^{-1}$, the central component is detected with a signal-to-noise ratio of $\sim$ 15. This brightest component is the only one detected in the spectrum of $^{13}$CO with an signal-to-noise ratio of $\sim$ 13. As a result, as shown in Fig.~\ref{fig:VLSR_pos_HI}, the left and right components are $^{12}$CO gases with different velocities, respectively. The spectrum of HI has a complex structure and several velocity line components; a Gaussian curve with five components was fitted to this spectrum. The brightest line (main line) and the next one to the right cover the three lines observed in $^{12}$CO and the main line in $^{13}$CO. We can explain this behaviour by the fact that these two HI lines are one spectral line that suffers self-absorption \citep[e.g.][]{Jackson2002}. All fitting parameters are shown in Tab.~\ref{table:adj1}.

After discovering [FKT-MC]2022 at V$_{\rm LSR} \sim -3$ km s$^{-1}$, and calculating its angular extension, we estimate its physical parameters, confirming through modeling the amount of energy required to produce the gamma-ray emission detected by LHAASO. Finally, both clouds, MML and [FKT-MC]2022, are compared to one another.  

\subsubsection{[FKT-MC]2022 Physical parameters}
\label{sec:size}

\begin{table*}[!ht]
	\caption{Fitted parameters of HI, $^{12}$CO(J=2$\rightarrow$1) and $^{13}$CO(J=2$\rightarrow$1) emission via Gaussian fit. The main beam (MB) averaged peak temperature ($T^p_{\rm MB}$) uncertainties are only due to rms noise. Velocity channel resolutions are used to show the LSR velocity (V$_{\rm LSR}$) and FWHM ($\Delta V$) uncertainties.}
	\centering
	\begin{tabular}{cccccc} 
		\hline
		Spectral & Size & V$_{\rm LSR}$ & $\Delta V$ & $T^{\rm p}_{\rm MB}$ & $\int T_{\rm MB} dv$ \\
		Line & [deg] & $\rm km \ s^{-1}$ & $\rm km \ s^{-1}$ & [K] & [$\rm K \ km \ s^{-1}$] \\

\hline

$^{12}$CO 2-1 (left) & 1.1 $\pm$ 0.2 & --12.19 $\pm$ 0.08 & 7.55 $\pm$ 0.19 & 0.43 $\pm$ 0.01 & 3.45 $\pm$ 0.10 \\
$^{12}$CO 2-1 center (main) & 1.1 $\pm$ 0.2 & --2.98 $\pm$ 0.01 & 5.21 $\pm$ 0.02 & 2.80 $\pm$ 0.01 & 15.54 $\pm$ 0.08 \\
$^{12}$CO 2-1 (right) &  1.1 $\pm$ 0.2 & 7.84 $\pm$ 0.03 & 3.78 $\pm$ 0.06 & 0.72 $\pm$ 0.01 & 2.88 $\pm$ 0.06 \\

\hline
\hline

$^{13}$CO 2-1 (main) &  1.0 $\pm$ 0.2 & --2.91 $\pm$ 0.03 & 3.31 $\pm$ 0.07 & 0.60 $\pm$ 0.01 & 2.12 $\pm$ 0.06 \\

\hline
\hline

HI (main) & 1.1 $\pm$ 0.2 & --10.18 $\pm$ 0.11 & 15.01 $\pm$ 0.25 & 108.46 $\pm$ 0.90 & 1740.37 $\pm$ 32.21 \\
HI (right to main) & 1.1 $\pm$ 0.2 & 4.42 $\pm$ 0.15 & 9.76 $\pm$ 0.30 & 58.62 $\pm$ 1.32 & 608.88 $\pm$ 23.34 \\

\hline
	\end{tabular}
	\label{table:adj1}
\end{table*}

In Tab.~\ref{table:adj2} we present the relevant parameters of the $^{12,13}$CO and HI fitted spectra (see Fig.~\ref{fig:spectra_2deg}) used to estimate the physical parameters of [FKT-MC]2022. Based on a gas region of 1.1$^{\circ}$ size (red ellipse in Fig.~\ref{fig:VLSR_pos_HI}), we examine only the central component of the $^{12}$CO(J=2$\to$1) spectrum and the single component fit for the $^{13}$CO(J=2$\to$1). In the case of the HI emission, we focus on the V$_{\rm LSR}$ range similar to the $^{12,13}$CO(J=2$\to$1) emission (between $\sim$ --4 and 9 km s$^{-1}$), which corresponds to the velocity range considered only in the main-beam averaged brightness temperature (see Fig.~\ref{fig:spectra_2deg}) 

We estimate T$_{\rm ex}$ from Eq. \ref{eq:Tex} using the $^{12}$CO(J=2$\to$1) emission line, assuming it is optically thick. The result with a mean value of $\approx$ 7.0 K is shown in Tab.~\ref{table:adj2}. This value agrees with the lower value given by \citet{Schneider2006} for several clumps in Cygnus-X (7-28 K) and is similar to the temperature (10 K) given by \citet{Dobashi2014} for the core L1004E of the dark nebula LDN 1004 in the COB7- MC.

Considering the local thermodynamic equilibrium (LTE) conditions for the $^{12,13}$CO(J=2$\to$1) emission, assuming an optically thin emission for $^{13}$CO, and using the main-beam peak brightness temperature of the fitted spectra (see Tab.~\ref{table:adj2}), we calculate the optical depth $\tau^{13}_{\rm CO}$ from the $^{13}$CO emission according to Eq. \ref{eq:tau}. Assuming a standard value for the isotopic abundance ratio of $\rm ^{12} CO / ^{13} CO \approx 60$ for the local interstellar medium \citep{Langer1993}, we estimate the optical depth of $^{12}$CO as $\tau^{12}_{CO}$ = 60 $\tau^{13}_{\rm CO }$. The estimated optical depths $\tau^{13}_{\rm CO }$ and $\tau^{12}_{\rm CO }$ can be found in Tab. \ref{table:adj2}.

We calculate the column densities of $^{12,13}$CO using Eqs.~\ref{eq:13CO_column} and \ref{eq:12CO_column} introducing the corresponding values presented in Tab.~\ref{table:adj2}. The HI column density is determined using Eq.~\ref{eq:eqnNHI} assuming optically thin emission. We show the estimated column densities for all emissions in Tab.~\ref{table:column1}. We obtain a mean value of N($^{\rm 13}$CO) $\approx$ 1.8 $\times$ 10$^{15}$ cm$^{-2}$, similar to the lower values reported by \citet{Schneider2006} for Cygnus-X (0.6 -- 6.9 $\times 10^{16}$ cm$^{-2}$). The H$_2$ column density is calculated from $^{12,13}$CO using Eq.~\ref{eq:H2_13CO_2}. The results are shown in Tab.~\ref{table:column1}. Mean values of N(${\rm H_2}$) = 3.1 $\times$ 10$^{21}$ and 9.1 $\times$ 10$^{20}$ cm$^{-2}$ are obtained from $^{12}$CO(J=2$\to$1) and $^{13}$CO(J=2$\to$1) emissions, respectively. These values agree in order of magnitude with those reported for Cygnus-X \citep{Uyaniker2001}. Finally, we estimate the total proton column density $N(\rm H)$ using both $^{12,13}$CO(J=2$\to$1) emission (separately) and Eq.~\ref{eq:eqnNH}. The calculated values are shown in Tab.~\ref{table:column2}. Nucleon column densities of $\rm N(H)$ =  $3.7 \times 10^{21}$ or $8.1 \times 10^{21}$ cm$^{-2}$ are obtained with $^{12}$CO(J=2$\to$1) or $^{13}$CO(J=2$\to$1) emission, respectively. 

The distance of 3.3 kpc to MML molecular cloud candidate \citep[][]{Cao2021b} was estimated using the rotation curve of \cite{Brand1993} by \cite{Miville2017}. For comparison, using the $^{13}$CO(J=2$\to$1) V$_{\rm LSR}$ of -2.9 km s$^{-1}$, we estimate a (far) kinematic distance of 1.7$^{+0.5}_{-0.7}$ kpc to [FKT-MC]2022 using the same rotation curve. The latter implies a physical size of the cloud of 33.4 $\pm$ 11.5 pc using a small angle approximation with an angular size of 1.1$^{\circ}$. Substituting the size of the cloud into Eq.~\ref{eq:eqnnH}, we calculate the corresponding n(H) via $^{12,13}$CO emission using both distances (3.3 and 1.7 kpc). We present the estimated n(H) values in Tab.~\ref{table:column2} using both distances to MML and [FKT-MC]2022. Using $^{13}$CO(J=2$\to$1) and $^{12}$CO(J=2$\to$1) emissions we obtain nucleon number densities of $\rm n(H) \sim$ 37 and $\sim$ 80  cm$^{-3}$, respectively. In addition, using the rotation curve of \cite{Reid2014},  we estimate a kinematic distance of $\sim$ 1.2$\pm$0.6 kpc, and a nucleon density of $\sim$ 52$\pm$28 $\rm cm^{-3}$ (via $^{13}$CO emission), comparable within uncertainties to that using a distance of 1.7 kpc.

Finally, for [FKT-MC]2022, we estimated the average number density of molecular hydrogen n$({\rm H_2})$ for both $^{12,13}$CO emissions using the Eq.~\ref{eq:eqnnH} using the estimated column density of H$_2$ in Tab.~\ref{table:column1}. The total mass of the molecular gas M($\rm H_2$) is obtained as follows:

\begin{equation}
\label{eq:H2mass}
M ({\rm H_2}) = \frac43 \pi (2\mu m_H) R^3 n({\rm H_2}),
\end{equation}

\noindent where $m_H =$ 1.67 $\times 10^{-24}$ g is the mass of the hydrogen atom and a molecular weight $\mu =$ 1.36 \citep{Miville2017,Takekoshi2019} is considered. We estimate the virial mass $\rm M_{vir}$ of the molecular cloud with \citep{Garay1999} as:

\begin{equation}
\label{eq:mass_virial}
    M_{\rm vir} = 210 \left[ \frac{R}{\rm pc} \right] \left[ \frac{(\Delta \rm V)}{\rm km s^{-1}} \right]^2,
\end{equation}

\noindent where we use the FWHM velocity width $\Delta \rm V$ of the $^{\rm 12,13}$CO fitted spectra from Tab~\ref{table:adj2}. The calculated $\rm H_2$ number densities and masses are shown in Tab.~\ref{table:mass}. Comparing this mass to the virial mass, it appears that [FKT-MC]2022 is close to virial equilibrium. The mean value of n(${\rm H_2}$) $\sim$ 20 cm$^{-3}$ for both $^{12,13}$CO emissions is similar to the mean value of $\sim$ 24 cm$^{-3}$ given in the $^{12}$CO(J=1$\rightarrow$0) catalogue of molecular clouds by \citet{Miville2017}, including MML. For MML using Eq. 23 of these authors: n(${\rm H_2}$) = 12 cm$^{-3}$; (see Tab.~\ref{table:mass}), which implies a density of nucleons (without HI) n(H) = 2n(${\rm H_2}$) = 24 cm$^{-3}$.

\begin{table*}
	\caption{\centering Values used for the column density analysis.}
	\centering
	\begin{tabular}{cccccc} 
		\hline
		Species & Diameter & $T_{\rm MB}^{\rm peak}$ & $\int T_{\rm MB} dV$ & $T_{\rm ex}$ & \textbf{$\tau^{12,13}_{\rm CO}$}  \\
		& [deg] & [K] &[K km s$^{-1}$] & [K] &    \\
		\hline

HI & 1.1 $\pm$ 0.2 & -- & 1018.45 $\pm$ 21.63 & -- & -- \\

$^{12}$CO 2-1 & 1.1 $\pm$ 0.2 & 2.80 $\pm$ 0.04 & 15.54 $\pm$ 0.08 & 7.16 $\pm$ 0.05 & 13.91 $\pm$ 1.23 \\

$^{13}$CO 2-1 & 1.1 $\pm$ 0.2 & 0.60 $\pm$ 0.05 & 2.12 $\pm$ 0.06 & 7.16 $\pm$ 0.05 & 0.23 $\pm$ 0.02 \\

\hline
	\end{tabular}
	\label{table:adj2}
\end{table*}

\begin{table*}
	\caption{\centering Estimated column density (N) of HI, $^{12}$CO, $^{13}$CO, and $\rm H_2$ for [FKT-MC]2022. The 12 and 13 labels correspond to the respective CO emission used in the calculations.  
	}
	\centering
	\begin{tabular}{cccccc} 
		\hline
		Diameter & $N(\rm ^{13}CO)$ & $N(\rm ^{12}CO)$ & $N(\rm HI)$ & $N^{\rm ^{12}CO}(\rm H_2)$ & $N^{\rm ^{13}CO}(\rm H_2)$ \\
		deg & [$10^{15}$ cm$^{-2}$] & [$10^{17}$ cm$^{-2}$] & [$10^{21}$ cm$^{-2}$] & [$10^{21}$ cm$^{-2}$] & [$10^{20}$cm$^{-2}$] \\
		\hline

		1.1 $\pm$ 0.2 & 1.8 $\pm$ 0.1 & 1.7 $\pm$ 0.2 & 1.9 $\pm$ 0.4 & 3.1 $\pm$ 0.9 & 9.1 $\pm$ 2.7 \\
		\hline

	\end{tabular}
	\label{table:column1}
\end{table*}

\begin{table*}
	\caption{\centering Estimated nucleon (H = 2H$_2$ + HI) column (N) and numeric (n) densities of the molecular cloud [FKT-MC]2022 using distances of 1.7 kpc and 3.3 kpc. The 12 and 13 labels correspond to the respective CO emission used in the calculations.}
	\centering
	\begin{tabular}{cccccccc} 
		\hline
		Diameter & $N^{12}(\rm H)$ & $N^{13}(\rm H)$ & $n^{12}(\rm H) {\scriptstyle (1.7 kpc)}$ & $n^{13}(\rm H) {\scriptstyle (1.7 kpc)}$ & $n^{12}(\rm H) {\scriptstyle (3.3 kpc)}$ & $n^{13}(\rm H) {\scriptstyle (3.3 kpc)}$ \\
		$[\rm deg]$ & [$10^{21} \rm cm^{-2}$] & [$10^{21} \rm cm^{-2}$] & [$\rm cm^{-3}$] & [$\rm cm^{-3}$] & [$\rm cm^{-3}$] & [$\rm cm^{-3}$]  \\
		\hline

1.1 $\pm$ 0.2 & 8.1 $\pm$ 1.9 & 3.7 $\pm$ 0.7 & 80 $\pm$ 34 & 37 $\pm$ 14 & 42 $\pm$ 10 & 19 $\pm$ 3 \\

\hline
	\end{tabular}
	\label{table:column2}
\end{table*}

\begin{table*}
	\caption{\centering Parameters of the molecular cloud [FKT-MC] 2022 and comparing with [MML2017]4607  }
	\centering
	\begin{tabular}{ccccccccc} 
		\hline
		Source & V$_{\rm LSR}$  & $D$ & Distance & $n({\rm H_2})$& $M_{\rm vir}({\rm H_2})$ & $M({\rm H_2})$  & $M(\rm HI + H_2)$ & Projected Size  \\
		& [$\rm km \ s^{-1}$]  & [deg] & [kpc] & [$\rm cm^{-3}$] & [$10^4 M_\odot$] & [$10^4 M_\odot$] & [$10^4 M_\odot$] & [deg] \\
\hline
MML & -13.71 & 0.5 & 3.3 & 12.26 & -- & 0.84 & -- & 0.5 at 3.3 kpc \\
FKT-MC ($^{12}$CO) & -3.0 $\pm$ 0.1  & 1.1 $\pm$ 0.2 & 1.7 $\pm$ 0.6 & 31 $\pm$ 14 & 9.5 $\pm$ 0.1 & 3.0 $\pm$ 1.4 & 3.6 $\pm$ 1.6 &  1.1 at 1.7 kpc \\

FKT-MC ($^{13}$CO) & -2.9 $\pm$ 0.1  & 1.1 $\pm$ 0.2 & 1.7 $\pm$ 0.6 & 9 $\pm$ 4 & 3.8 $\pm$ 0.2  & 0.9 $\pm$ 0.4 & 1.5 $\pm$ 0.6 & 1.1 at 1.7 kpc \\

\hline

	\end{tabular}
	\label{table:mass}
\end{table*}

\subsubsection{Modelling gamma-rays}
\label{sec:gamma_res}

\begin{figure}[ht!]
\includegraphics[width =\columnwidth]{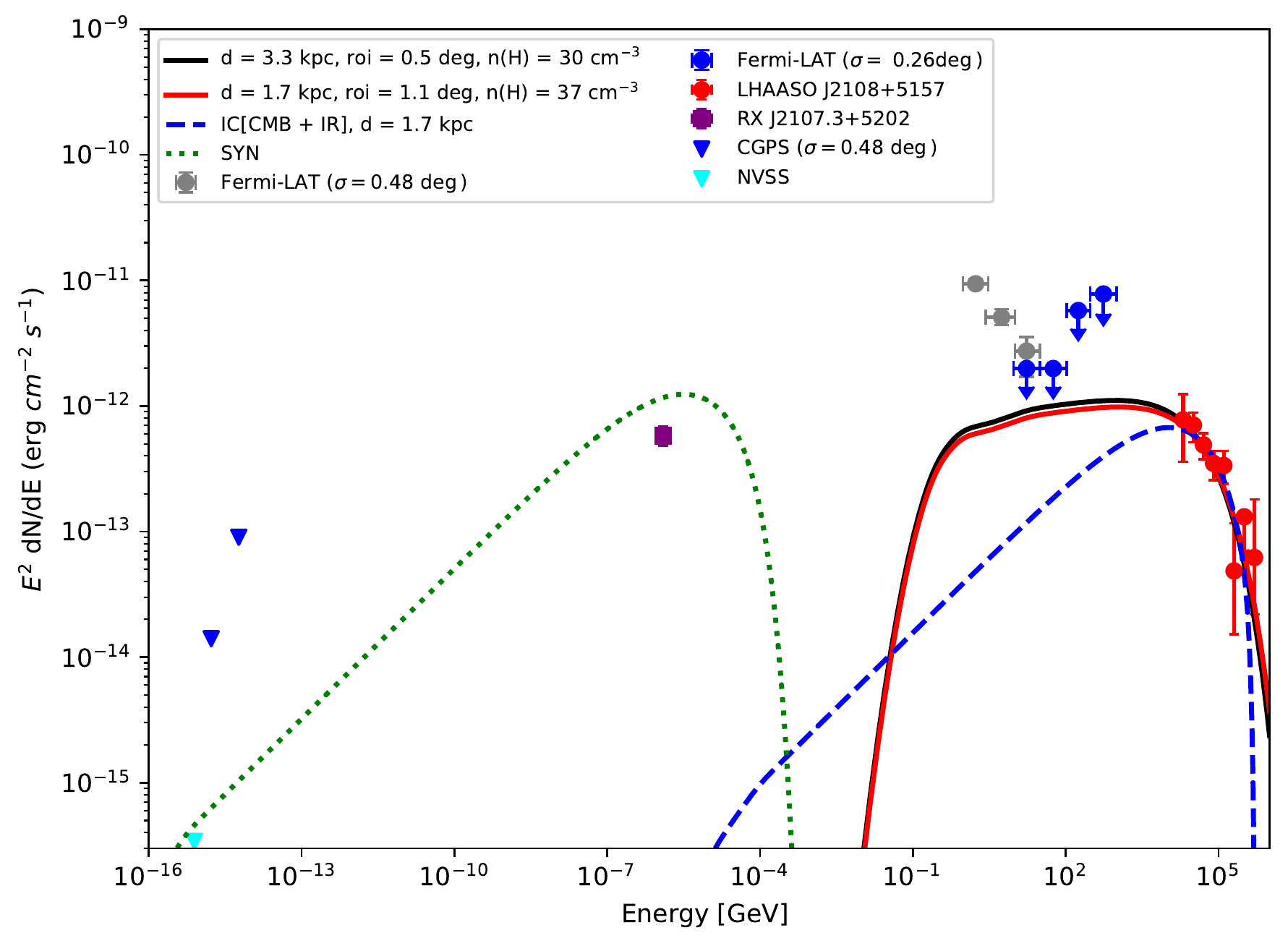}

\caption{Spectral energy distribution of the LHAASO J2108+5157 region. Red circles are from the LHAASO observatory \citep{Cao2021b}. The grey and blue circles are GeV fluxes and the upper limit reported by the Fermi-LAT observatory, respectively \citep[][and references therein]{Abdollahi2020}. The purple square is the ROSAT X-ray source RX J2107.3+5202 \citep{Motch1997}. The blue and cyan triangles are radio observations from CGPS survey \citep{Taylor2003} and NVSS survey \citep{Condon1998} respectively. The green dotted curve is the spectral energy distribution expected from synchrotron radiation. The blue dashed curve represents the expected spectral energy from the inverse Compton effect, considering a CMB and IR seed photon field (see text for details). The solid curves are the expected gamma-ray spectra for MML (black) and [FKT-MC]2022 (red), respectively. The SED parameters for hadronic models are shown in Tab.~\ref{tab:hadron}.}

\label{fig:SED_hadronic_leptonic}
\end{figure}

\begin{table*}
    \caption{\centering Results of hadronic models of Naima, considering Results for MML and [FKT-MC]2002}
    \centering
    \begin{tabular}{lcccccccc} \hline
        & Distance & n(H$_2$) & n(H) & ROI & $W_p$ & Cutoff & Molecular & Remark \\ 
         & [kpc] & [cm$^{-3}$] & [cm$^{-3}$] & [degree] & [$10^{47}$erg] & [TeV] & Observation \\ \hline
      MML   & 3.3 & --- & 30 & 0.5 & 9$_{-2}^{+4}$ & $600_{-300}^{+300}$ & $^{12}$CO(J=1$\to$0) & n(H); \cite{Cao2021b} \\
      MML   & 3.3 & 12 & --- & 0.5 & 22 $\pm$ 7 & $700_{-300}^{+400}$ & $^{12}$CO(J=1$\to$0) & n(H$_2$); Miville et al. (2017)\\
      MML    & 3.3 & --- & 24 & 0.5 & 12 $\pm$ 3 & $600_{-200}^{+300}$ & $^{12}$CO(J=1$\to$0) & n(H)=2n(H$_2$)\\
      $[{\rm FKT}$-${\rm MC}]$ 2022 & 1.7 & 9 & --- & 1.1 & 7 $\pm$ 2 & $700_{-300}^{+400}$ & $^{13}$CO(J=2$\to$1) \\
      $[{\rm FKT}$-${\rm MC}]$ 2022 & 1.7 & ---& 37 & 1.1 & 1.7$_{-0.4}^{+0.7}$ & $700_{-300}^{+400}$ & $^{13}$CO(J=2$\to$1) \\
      \hline

    \end{tabular}
    \label{tab:hadron}     

\end{table*}

\begin{figure}[ht!]

\begin{center}
\includegraphics[width=\columnwidth]{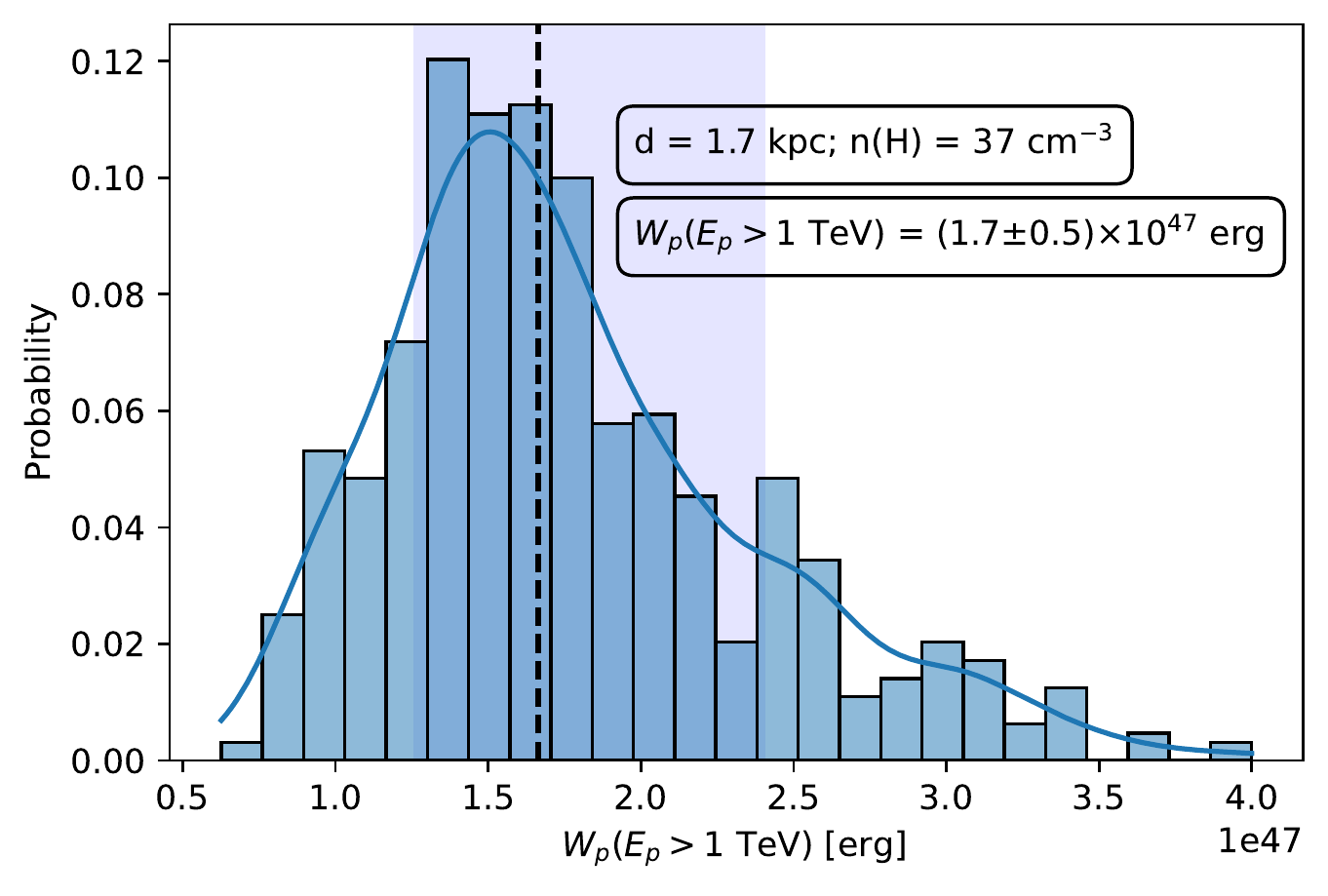}
\includegraphics[width=\columnwidth]{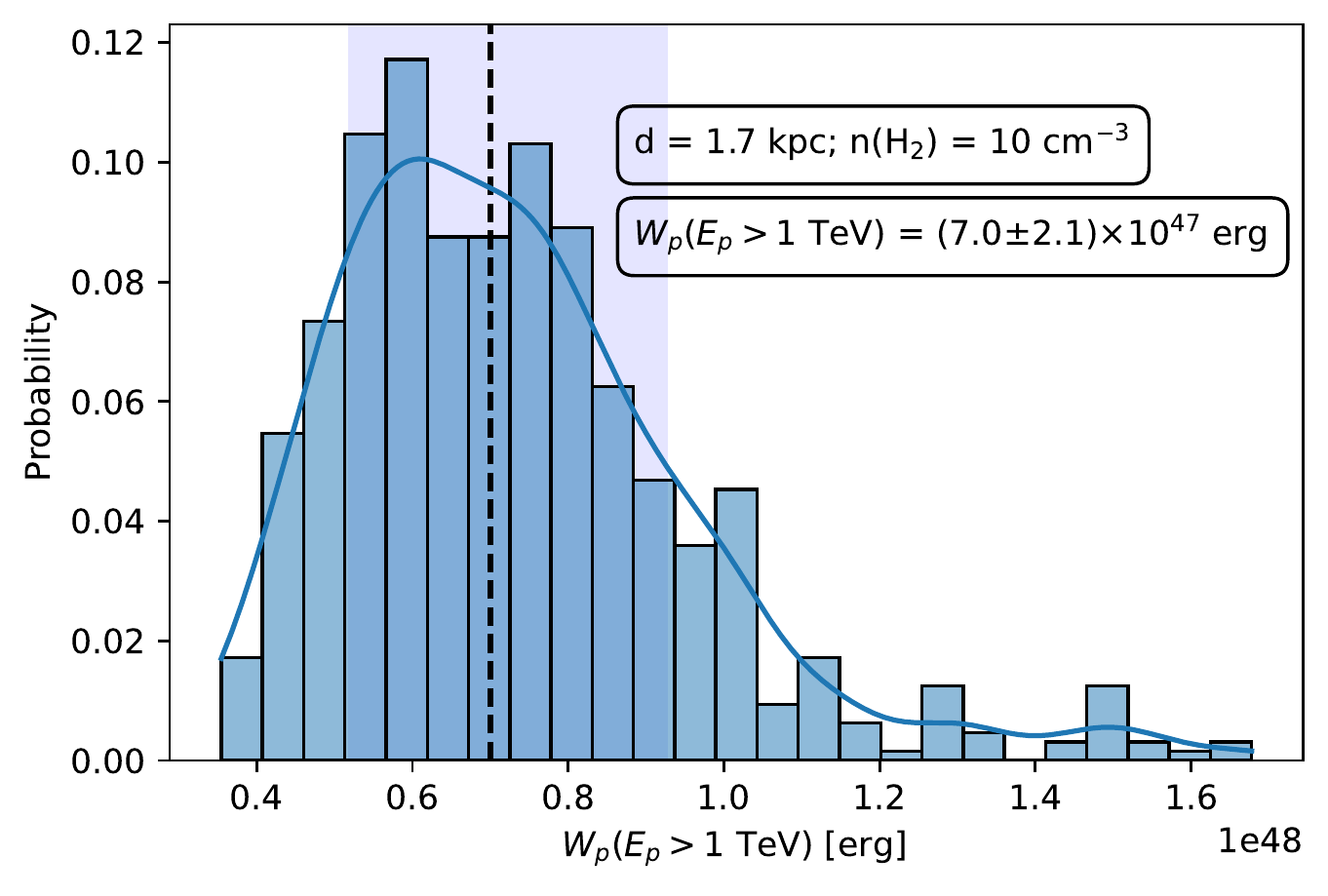}
\includegraphics[width=\columnwidth]{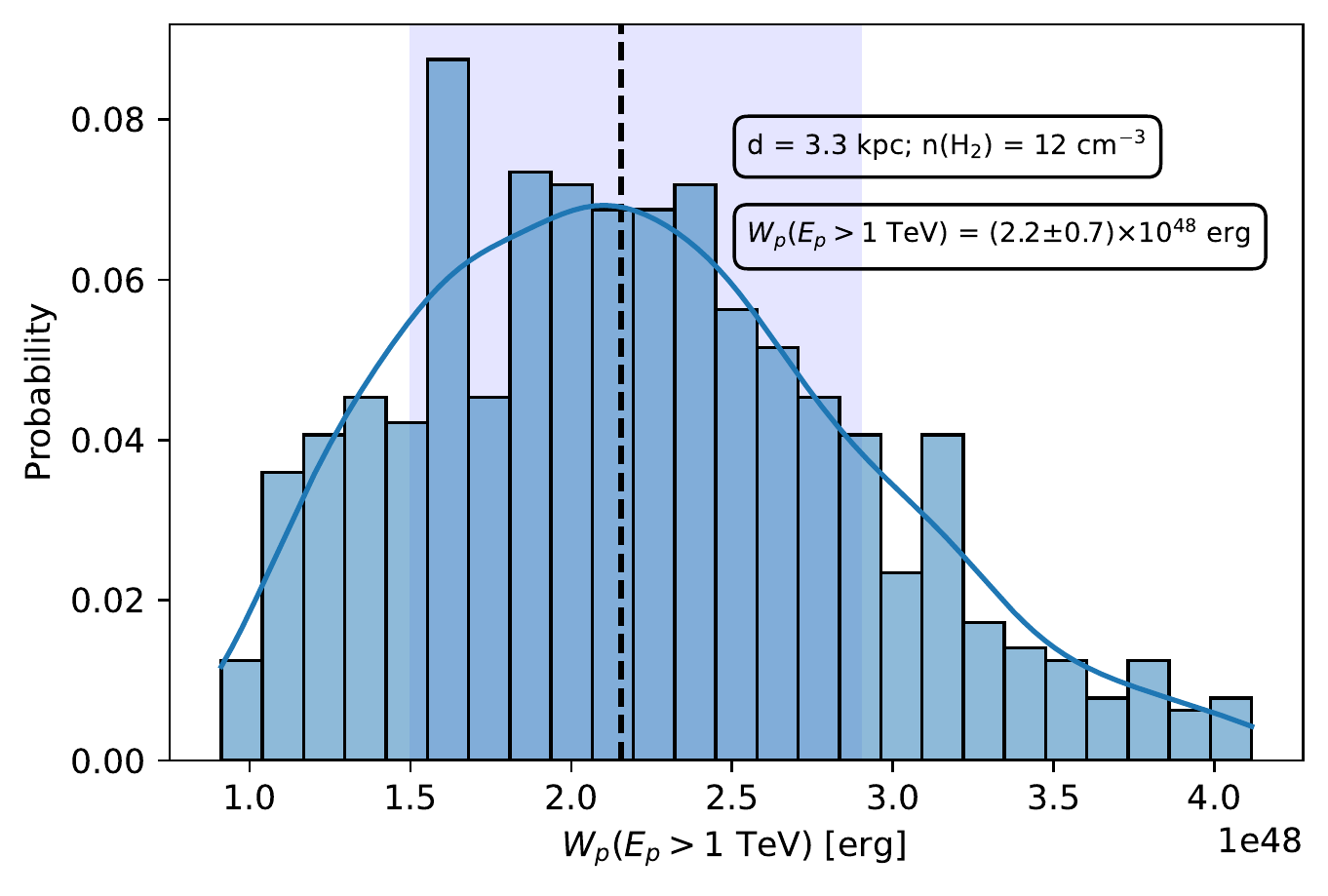}
\end{center}
 \caption{Computed Hadronic model using Naima for [FKT-MC]2022, considering LHAASO parameters \citep{Cao2021a}, a distance of 1.7 kpc, and optically thin gas for a) density of nucleons n(H) = 37 cm$^{-3}$, and b) n(H$_2$) = 10 cm$^{-3}$ (this work). c) The same for the [MML2017]4607 molecular cloud, but considering a distance of 3.3 kpc, and optically thick gas with n(H$_2$) = 12 cm$^{-3}$ from \citet{Miville2017}. }
    \label{fig:naima}
\end{figure}

The first spectral energy distribution (SED) of J2108 was presented by \citet{Cao2021b}. Nevertheless, this SED does not include RXJ2107.3$+$5202 and the NVSS emission. In Fig.~\ref{fig:SED_hadronic_leptonic} we show this SED with these two emissions. The SED was calculated using the models of \citet{Zabalza2016} in Naima software\footnote{url{https://naima.readthedocs.io/en/latest/index.html}}. A similar complementary SED is presented by \citet{Kar2022}. At high energies, it is complicated to distinguish between the leptonic and hadronic nature of the emission (see \S~\ref{sec:J2108} and \citealt{Cao2021b}), although the leptonic nature seems to be favoured at lower energies.

The hadronic and leptonic modeling considers a power law spectrum with an exponential cutoff and fixed spectral indexes of 2.0 and 2.2, respectively. In the leptonic modeling we adopt an energy cutoff of 200 TeV and a magnetic field strength of 3 $\mu$G \citep{Cao2021b}. For IC scattering, we include CMB and IR seed photon fields with temperatures of 2.72 and 20.0 K, and energy densities of 0.26 and 0.3 eV cm$^{-3}$, respectively \citep[e.g.][]{Hinton2009, Vernetto2016}. Considering a distance of 1.7 kpc, we obtain a total energy of electrons with energy above 1 GeV of 2.2 $\pm$ 0.4 $\times 10^{46}$ erg. Nevertheless, no clear source for the leptonic emission nor extended X-ray emission in the vicinity (neither in the FOV nor in Kron 82) has been observed or reported.

Regarding this fact and the presence of two molecular clouds (MML and [FKT-MC]2022) as candidates for the generation of the sub-PeV emission, we favour the scenario of a hadronic origin: gamma rays generated by the decay of neutral pions due to the interaction between accelerated protons and molecular clouds. Here, the pion decay model in \citet{Zabalza2016} is based on \citet{Kelner2006,Kamae2006} and implemented by \citet{Kafexhiu2014}. We used a distance of 3.3 kpc for MML \citep{Cao2021b} and 1.7 kpc for [FKT-MC]2022, considering ROIs of 0.5$^{\circ}$ and 1.1$^{\circ}$, respectively. We use a nucleon number density of 30 cm$^{-3}$ \citep[estimated by][]{Cao2021b} for MML and 37 cm$^{-3}$ for [FKT-MC]2022, and calculate the total proton energy as follows:

\begin{equation}
\label{eq:eqnEP33}
\Bigg[\frac{\rm E_{P}}{\rm erg}\Bigg]=9.0\pm 3.0\times 10^{47}\Bigg[\rm \frac{n(H)}{30~cm^{-3}}\Bigg]^{-1} \Bigg[\rm \frac{D}{3.3~kpc}\Bigg]^{2} \quad \rm{for \ MML},
\end{equation}

and

\begin{equation}
\label{eq:eqnEP17}
\Bigg[\rm \frac{ E_{P}}{erg}\Bigg] =  1.7\pm 0.5\times 10^{47}\Bigg[\rm \frac{n(H)}{37~cm^{-3}}\Bigg]^{-1} \Bigg[\rm\frac{D}{1.7~kpc}\Bigg]^{2} \\  \rm{for\ FKT}.
\end{equation}

In Tab.~\ref{tab:hadron}, we show the results of several hadronic models (via neutral pion decay process) by using Naima software, considering different parameters for [FKT-MC]2022 and MML. In Fig.~\ref{fig:naima} we show the energy distribution of the proton population (with energy above 1 TeV) of three of the hadronic models: [FKT-MC]2022 with a proton density of 37 cm$^{-3}$, [FKT-MC]2022 with an optically thin H$_2$ density of 10 cm$^{-3}$, and MML with an optically thick H$_2$ density of 12 cm$^{-3}$ \citep{Miville2017}. Comparing these energy distributions, the W$_p$ (7$\pm$2$\times$10$^{47}$ erg) of [FKT-MC]2022 (optically thin gas with H$_2$ density of 10 cm$^{-3}$) is close to that for MML of 2.2$\pm$0.7$\times$10$^{48}$ erg (optically thick gas with H$_2$ density = 12 cm$^{-3}$). The latter is consistent with the W$_p$ = 2.0$\times$10$^{48}$ erg reported in \citet{Cao2021b}, which was calculated using the Naima software with a distance of 3.3 kpc, and optically thick gas approximation with n(H) = 2n(H$_2$) =  30 cm$^{-3}$. 

On the other hand, for [FKT-MC]2022 at 100 TeV and assuming no leptonic contribution, we obtain a flux of E$^2$ $\frac{dN}{dE}$ $\sim$ 2 $\times$ 10$^{-13}$ erg cm$^{-2}$ s$^{-1}$ from the hadronic contribution of Fig.~\ref{fig:SED_hadronic_leptonic}. The latter corresponds to a differential flux of N$_\gamma^{\rm had}$ = 1.25 $\times$ 10$^{-17}$ photons TeV$^{-1}$ cm$^{-2}$ s$^{-1}$. This value corresponds to the N$_P({\rm CR })$ = 1.04 $\times$ 10$^{-29}$ TeV$^{-1}$ in Eq.~\ref{eq:eqnmodel} 

[FKT-MC]2022 (see Eq. ~\ref{eq:eqnEP17}) reproduces an observed differential flux of $\sim$ 10$^{-17}$ photons TeV$^{-1}$ cm$^{-2}$ s$^{-1}$ (Fig.~\ref{fig:SED_hadronic_leptonic}) assuming an optically thin nucleon density of n(H) = 37 cm$^{-3}$. This flux corresponds to an E$_{\rm P}$ ($\sim 1.7 \times 10^{47}$ erg), which is slightly less than the required energy for MML ($\sim 9.0 \times 10^{47}$ erg; optically thick gas). In addition, a lower required energy of $E_{\rm P} =$ 6.7 $\pm$ 2.3 $\times 10^{46}$ ergs is obtained [FKT-MC]2022, if a distance of 1.2 kpc (obtained with the rotation curve of \citealt{Reid2014}) and a nucleon density of 52 cm$^{-3}$ are considered. Finally, if we project the size of [FKT-MC]2022 to the distance of MML (3.3 kpc) or if we project MML to a distance of [FKT-MC]2022, the size of the clouds matches accordingly (see Tab.~\ref{table:mass}).

The energies of both molecular clouds are much lower than the energy of a single SN ($\sim$ 10$^{51}$ erg). Therefore, an ancient unidentified SNR could be the PeVatron\citep[e.g][and references therein]{Kar2022}. Ultimately, we favour the presence of a molecular cloud, MML or [FKT-MC]2022, as a place to produce the emission observed by LHAASO in J2108. High-resolution radio observations, including denser gas tracers, are needed to clarify the scenery.

\section{Conclusions}
\label{sec:conclusion}

We present low-resolution $^{12,13}$CO(2$\rightarrow$1) observations made with the 1.85-m radio-telescope at Osaka Prefecture University to detect a molecular counterpart to the PeVatron associated with LHAASO J2108+5157 in the OB7 molecular cloud at Cygnus. We discuss the methodology to obtain the observed flux of gamma-ray emission by determining an appropriate density of nucleons, including HI and H$_{2}$ (through CO observations) emissions:

\begin{itemize}

\item In addition to MML[2017]4607, we propose molecular cloud [FKT-MC]2022 as another candidate to produce the observed emission for LHAASO J2108+5157, favouring the hadronic emission because neither a source nor data support leptonic emission.

\item $[$FKT-MC$]$2022 is situated at a distance of 1.7 $\pm$ 0.6 kpc. It is $\sim$ 1.1$^{\circ}$ in size and has nucleon densities (HI + H$_2$) of $\sim$ 80 and 37 cm$^{-3}$ for $^{12}$CO (optically thick) and $^{13}$CO (optically thin) emission respectively. These values correspond to M(HI +H$_2$) of $\sim$ 4$\times$10$^4$ M$_{\odot}$ and 2$\times$10$^4$ M$_{\odot}$ respectively. We computed a total required energy of protons of W$_p \sim 1.7 \times 10^{47}$ erg to reproduce the gamma-ray emission observed by LHAASO. 

\item The $^{12,13}$CO(2$\rightarrow$1) observations of Kronberger 82 reveal a clump morphology with a V$_{\rm LSR}$ $\sim$ --7 km s$^{-1}$ and a size of 0.1$^{\circ}$. For optically thin gas at a distance between 1.63 and 2.30 kpc, the n(HI+2H$_2$) or n(H) is between 1.7 and 2.5 $\times$ 10$^3$ cm$^{-3}$  and M(HI +H$_2$) is between 0.4 and 1.7 $\times$ 10$^3$ M$_{\odot}$. These values are consistent with the optically thick gas and with other molecular observations, confirming Kronberger 82 as a star-forming region rich in molecular species 

\item By modelling its hadronic gamma-ray emission, Kronberger 82 uses less energy (W$_p \sim 2.6-7.0 \times 10^{45}$ erg) to produce the observed (sub)PeV gamma-ray emission in LHAASO J2108+5157 than [FKT-MC]2022 and MML[2017]4607. Nevertheless, due to its angular separation from J2108+5157, it is not a strong candidate to be the gamma-ray source. \\

\end{itemize}

\noindent \textbf{\large Conflict of Interest}

\noindent The authors state that have no conflict of interest directly relevant to the content of this article

\begin{ack}

This work was (partially) supported by the Inter-University Research Program of Institute for Cosmic Ray Research (ICRR), the University of Tokyo. IT--J acknowledges support from Consejo Nacional de Ciencia y Tecnolog\'ia (CONACyT), M\'exico; grant 754851. We are grateful for the computational resources and technical support offered by the Data Analysis and Supercomputing Center (CADS) through the Leo-Atrox supercomputer of the Universidad de Guadalajara. EdelaF thanks colegio departamental, departamento de F\'isica, and respective authorities of the Centro Universitario de Ciencias Exactas e Ingenier\'ias (CUCEI), Universidad de Guadalajara, for authorization and permissions to perform the academic stays.

\end{ack}





\appendix

\renewcommand{\thechapter}{\Alph{chapter}}

\numberwithin{equation}{chapter}

\chapter{Determination of column densities and n(H)}
\label{appendix:ap1}

First, we compare the distribution of HI with the $^{12,13}$CO gas to reliably clarify the position of the associated molecular cloud. Then we obtain the spectra and calculate the physical parameters in a standard way.

Assuming optically thin emission, {the column density (N) using HI is computed as \citep{Dickey1990}:

\begin{equation}
\label{eq:HI_column_2}
    \left[ \frac{N ({\rm HI})}{\rm cm^{-2}} \right]= 1.823 \times 10^{18} \left[ \frac{\int T_{\rm MB}^{HI} dv}{\rm K \ km \ s^{-1}}\right],
\end{equation}

\noindent where $T_{\rm MB }^{ HI }$ is the averaged brightness temperature of the main beam (MB) resulting from a Gaussian fit of the spectrum of HI emission.

\noindent Under LTE conditions and assuming $^{12}$CO optically thick gas the excitation temperature $T_{\rm ex}$ is:

\begin{equation}
\label{eq:Tex}
    T_{\rm ex} = \frac{T_0}{\ln \left( 1 + \frac{T_0}{T_{\rm MB}^{12,p} + T_0 / \left(\exp ( T_0 / T_{\rm CMB})-1\right)} \right)},
\end{equation}

\noindent where $T_{\rm MB }^{12,p}$ is the main beam averaged peak temperature of the $^{12}$CO emission, T$_{\rm CMB}$ = 2.725 K is the temperature of the cosmic microwave background (CMB) and $T_0 = h \nu / k$ is the characteristic temperature of the $^{12,13}$CO (J=2$\rightarrow$1) at a rest frequency of $\nu^{12}_{\rm CO} =$ 230.538 and $\nu^{13}_{\rm CO} =$ 220.398 GHz, respectively. The optical depth $\tau^{13}_{\rm CO}$ of the $^{13}$CO gas is determined assuming an optically thin emission and using \citep{Rohlfs2004}: 

\begin{equation}
\label{eq:tau}
    \tau^{13}_{\rm CO} = - \ln \Bigg[ 1 - \frac{J_\nu^{12}(T_{\rm ex}) - J_\nu^{12}(T_{\rm CMB})}{J_\nu^{13}(T_{\rm ex}) - J_\nu^{13}(T_{\rm CMB})} \frac{T_{\rm MB}^{13,p}}{T_{\rm MB}^{12,p}} \Bigg],
\end{equation}

\noindent where $T_{\rm MB }^{13,p}$ is the main beam averaged peak temperature of the $^{13}$CO emission and,

\begin{equation}
J_\nu(T) = \frac{h \nu / k}{e^{h \nu / k T} - 1}, 
\end{equation}

\noindent is the intensity in units of temperature related with the CMB and the excitation temperature for the $^{12,13}$CO(J=2$\rightarrow$1) emission.

From \citet{Scoville1986} and \citet{Palau2007} we derived an expression to calculate the column density of $^{13}$CO from a rotational transition $J \to J - 1$ as

\begin{equation}
\label{eq:13CO_column}
    \begin{split}
    \left[ \frac{N(\rm ^{13}CO)}{\rm cm^{-2}}\right] &= 4.70 \times 10^{13} \frac{T_{\rm ex}}{J^2} \exp \left( \frac{2.64 \ J(J+1)}{T_{\rm ex}}\right) \\
& \times \frac{\tau^{13}_{\rm CO}}{1-e^{-\tau^{13}_{\rm CO}}} \left[ \frac{\int T_{\rm MB}^{13}(v) dv}{\rm K \ km \ s^{-1}}\right],
\end{split}
\end{equation}

\noindent where the integrated brightness temperature of the main beam for the $^{13}$CO emission, $\int T^{13}_{\rm MB } (v) dv$, is obtained from the Gaussian fit for the respective velocity component of the $^{13}$CO spectrum. Taking a standard value for the isotopic abundance ratio of $\rm ^{12} CO / ^{13} CO \approx 60$ for the ISM \citep{Langer1993}, we estimate the optical depth of $^{12}$CO as $\tau^{12}_{\rm CO} $= 60$\tau^{13}_{\rm CO}$. Similarly, the column density of $^{12}$CO can then be calculated with \citep{Palau2007}:

\begin{equation}
\label{eq:12CO_column}
\begin{split}
    \left[ \frac{N(\rm ^{12}CO)}{\rm cm^{-2}}\right] &= 4.33 \times 10^{13} \frac{T_{\rm ex}}{J^2} \exp \left( \frac{2.77 \ J(J+1)}{T_{\rm ex}}\right) \\
& \times \frac{\tau^{12}_{\rm CO}}{1-e^{-\tau^{12}_{\rm CO}}} \left[ \frac{\int T_{\rm MB}^{12}(v) dv}{\rm K \ km \ s^{-1}}\right].
\end{split}
\end{equation}

\noindent where $\int T^{12}_{\rm MB } (v) dv$ is the integrated brightness temperature of the main beam for the $^{12}$CO emission.} Using an abundance ratio of $[\rm H_2 / ^{13} CO] = 10^{5}$ \citep{Dickman1978,Pineda2008}, the column density of H$_2$ (via the $^{13}$CO emission) can be estimated by

\begin{equation}
\label{eq:H2_13CO_2}
N(\rm{H_2}) = 5 \times 10^5 \  N(^{13}{\rm CO})
\end{equation}

\noindent Moreover, several studies \citep[e.g][]{Bolatto2013} find a ``$^{12}$CO-to-H$_2$'' conversion factor with a Galactic mean of $X_{\rm CO } =$ 2.0 $\times 10^{20}$ cm$^{-2}$ (K km s$^{-1}$)$^{-1}$ with 30\% uncertainty within the Milky Way disk. Then, the H$_2$ column density (via the $^{12}$CO emission) can be estimated as:

\begin{equation}
\label{eq_H2_12CO}
N({\rm H_2}) = X_{\rm CO} \int T_{\rm MB}^{12}(v) dv
\end{equation}

\noindent Finally, the total proton column density N(H) is calculated using Eq.~\ref{eq:eqnNH}.

\chapter{IRAS 21046+5110 and Kronberger 82}
\label{appendix:ap3}

\textbf{B.1 IRAS 21046+5110}

IRAS 21046 is a star-forming region with more than 200 massive protostellar candidates in 54 embedded clusters at a distance of $\sim$ 600 pc \citep{Kumar2006}. These authors present a complete and detailed NIR study confirming that the clusters and their properties have more to do with the formation process than with evolution. Considering its stellar content mass, M$_{\rm stellar}$, of 17 M$_{\odot}$ \citep{Kumar2006}, the clear but fainter OPU $^{\rm 13}$CO(J=2$\to$1) emission compared to Kron 82 (integrated flux ratio $\sim$ 0.3; see Fig.~\ref{fig:cygOB7}), its farthest distance from J2021 compared to Kron 82 ($\sim$ 0.7$^{\circ}$ vs. $\sim$ 0.4$^{\circ}$), and the fact that its nature is not comparable to the Cygnus OB2, we do not support it as a producer of the sub-PeV emission observed by LHAASO. Furthermore, it was not considered in the inspection of \citet{Cao2021b}.

\

\textbf{B.2 Kronberger 82} 
\label{sec:Kron82}

Kron 82 is a poorly studied star-forming region. A visual inspection of IR using 2MASS (images not shown), IRAC, WISE, and NVSS emission (see lower panels in Fig.~\ref{fig:lhaaso_opt}) suggests that Kron 82 has the classic IR morphology observed in star-forming regions, including ultra-compact HII regions with extended emission and OB stars covered by arcminute-sized RC emission \citep[e.g][and references therein]{delaFuente2020a,delaFuente2020b, Churchwell2007}. The latter implies that ionised gas coexists with star clusters (2.12 $\mu$m, 3.6 $\mu$m, 4.5 $\mu$m and RC at 20 cm; in agreement with the corresponding WISE emission), with associated dust traced at WISE 22 $\mu$m (no IRAC data at 5.8, 8.0 and 24 $\mu$m). In this context, the remarkable HII region IRAS 21078+5211 (distance of $\sim$ 1.6 kpc; \citealt{Moscadelli2021} and references therein), the star-forming region IRAS 21078+5209 (distance of $\sim$ 1.5 kpc; see \cite{Gieser2021} and references therein), and the star cluster DSH J2109.5+5223 \citep{Kronberger2006} are located in the interior of the Kron 82 region. Nevertheless, a literature search for massive and WR stars in the vicinity, and a quick 2MASS photometry shows that Kron 82 resembles more IRAS 21046+5110 than Cygnus OB2: ruled by YSOs and no WR detected. A more detailed study of the stellar content in IR is needed to determine the nature of Kron 82 as a star cluster (formation dominated by YSOs like IRAS 21046, or evolution dominated like Cygnus OB2 with intense stellar winds).

\begin{figure*}[!ht]
\begin{center}
\includegraphics[width=0.9\textwidth]{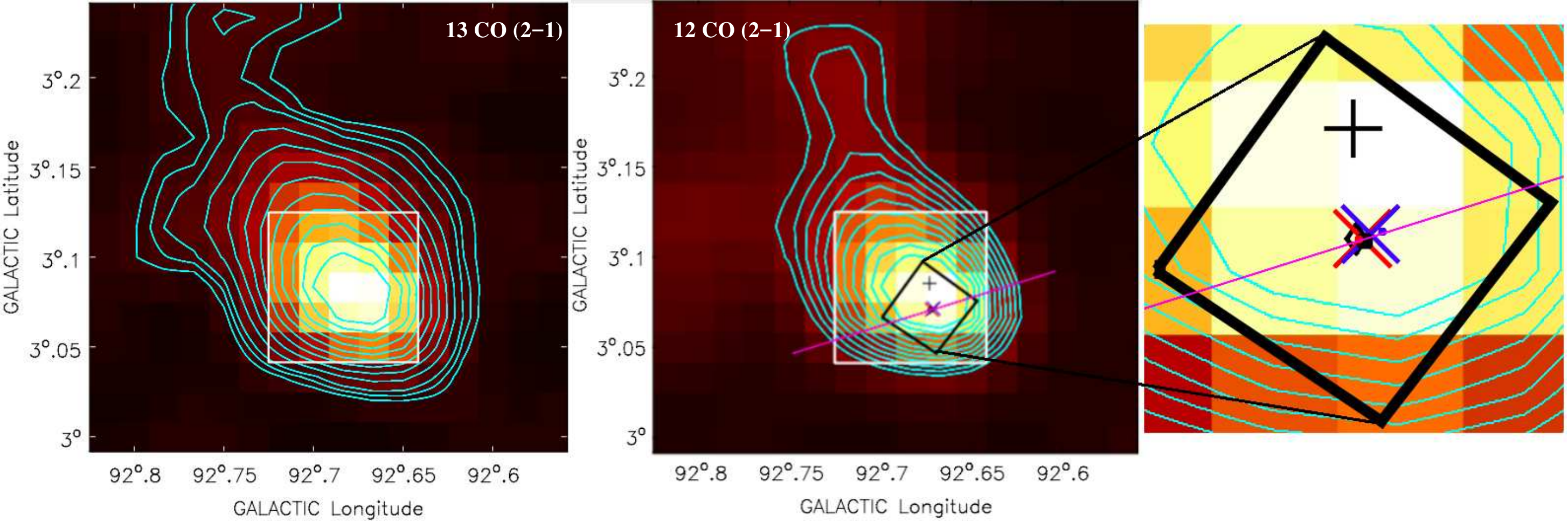}
\end{center}
\caption{Same as the bottom left of Fig.~\ref{fig:lhaaso_opt}, but for the OPU $^{12,13}$CO(J=2$\rightarrow$1) emission of Kron 82 (colours and contours). The selected region of 0.1$^{\circ}$ that was used to extract the spectrum in Fig.~\ref{fig:Spec_K82_CO} is shown as a square in the leftmost panel of the Figure. In the middle panel, the rotated square, which indicates the direction of the equatorial coordinates, shows the region studied at higher angular--resolution by \citet{Moscadelli2021}; see their Fig. 2. The peak position of the IRAM 30 m data is shown as a diamond mark right at the centre of the rotated square. In the same panel the cross corresponds to the tip of the protrusion of the emission observed their Fig. 2. The blue and red "x" marks indicate the position of the blue-and red--shifted, respectively, SW lobes originating from source 1 (see their Figs 1 and 5). The line shows the projection of these lobes when transferred to larger scales. The size of the beam is 0.045$^{\circ}\times$0.045$^{\circ}$..}
\label{fig:K82_COs}
\end{figure*}

\begin{figure*}[!ht]
\begin{center}
\includegraphics[width=\columnwidth]{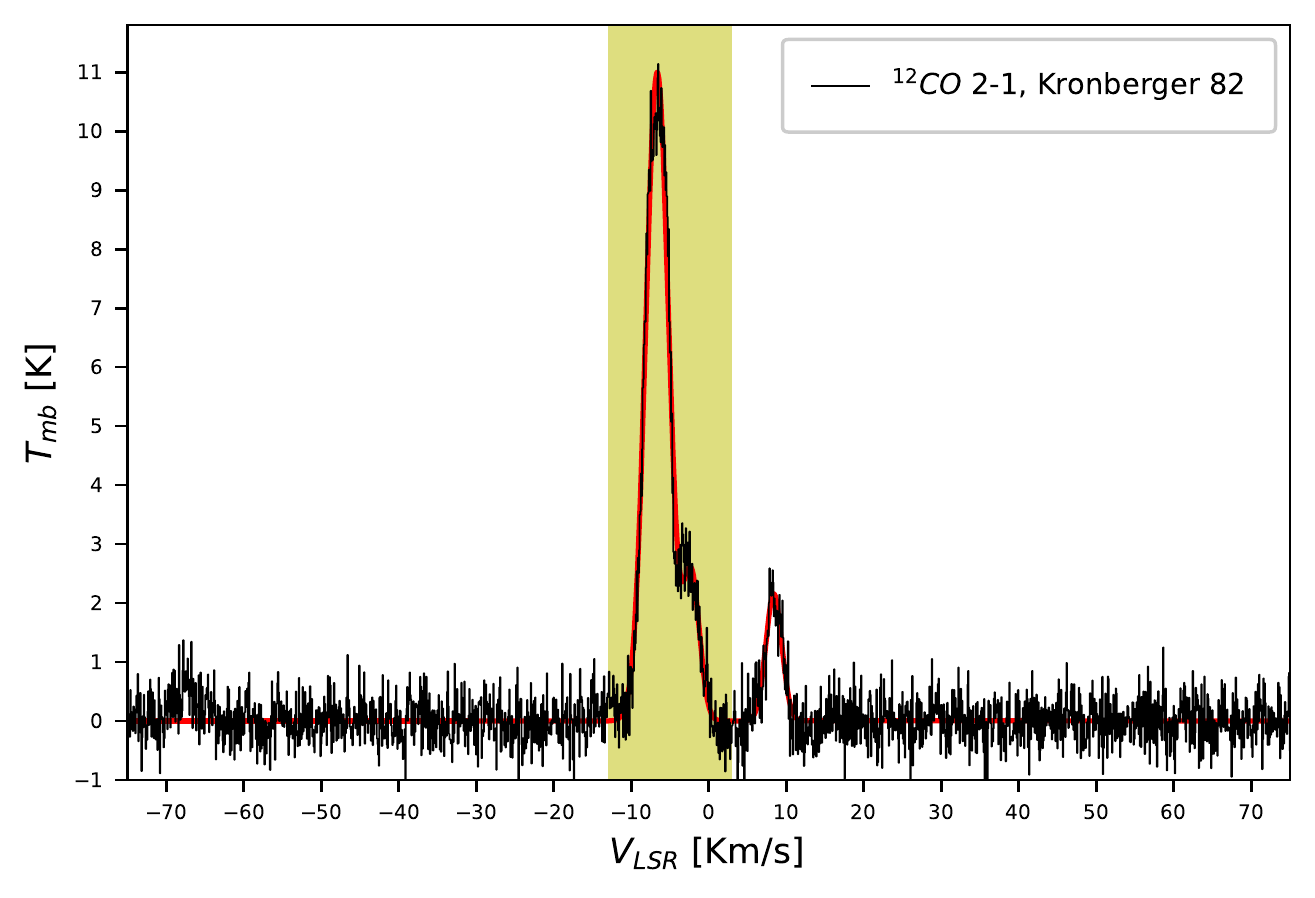}
\includegraphics[width=\columnwidth]{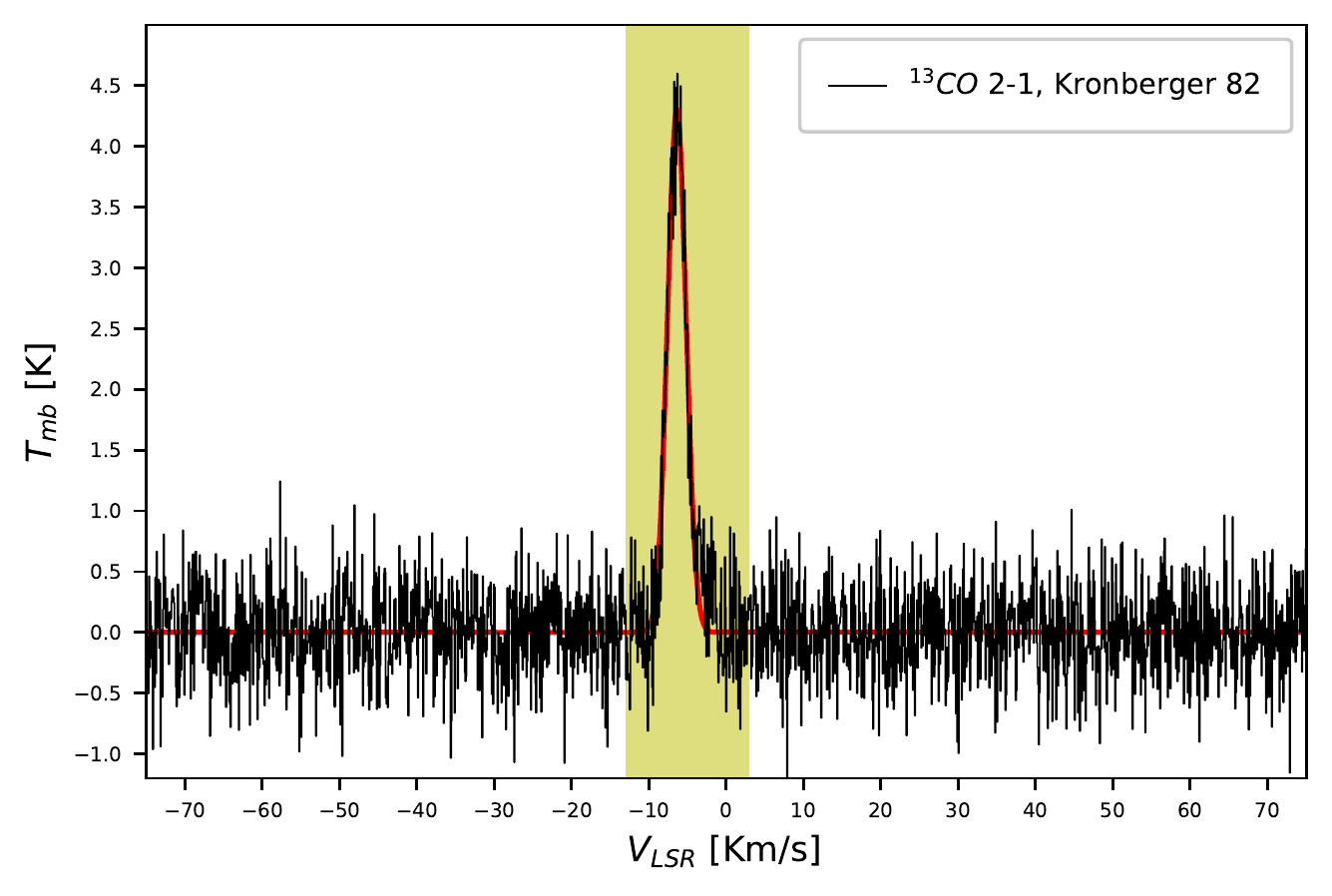}
\includegraphics[width=\columnwidth]{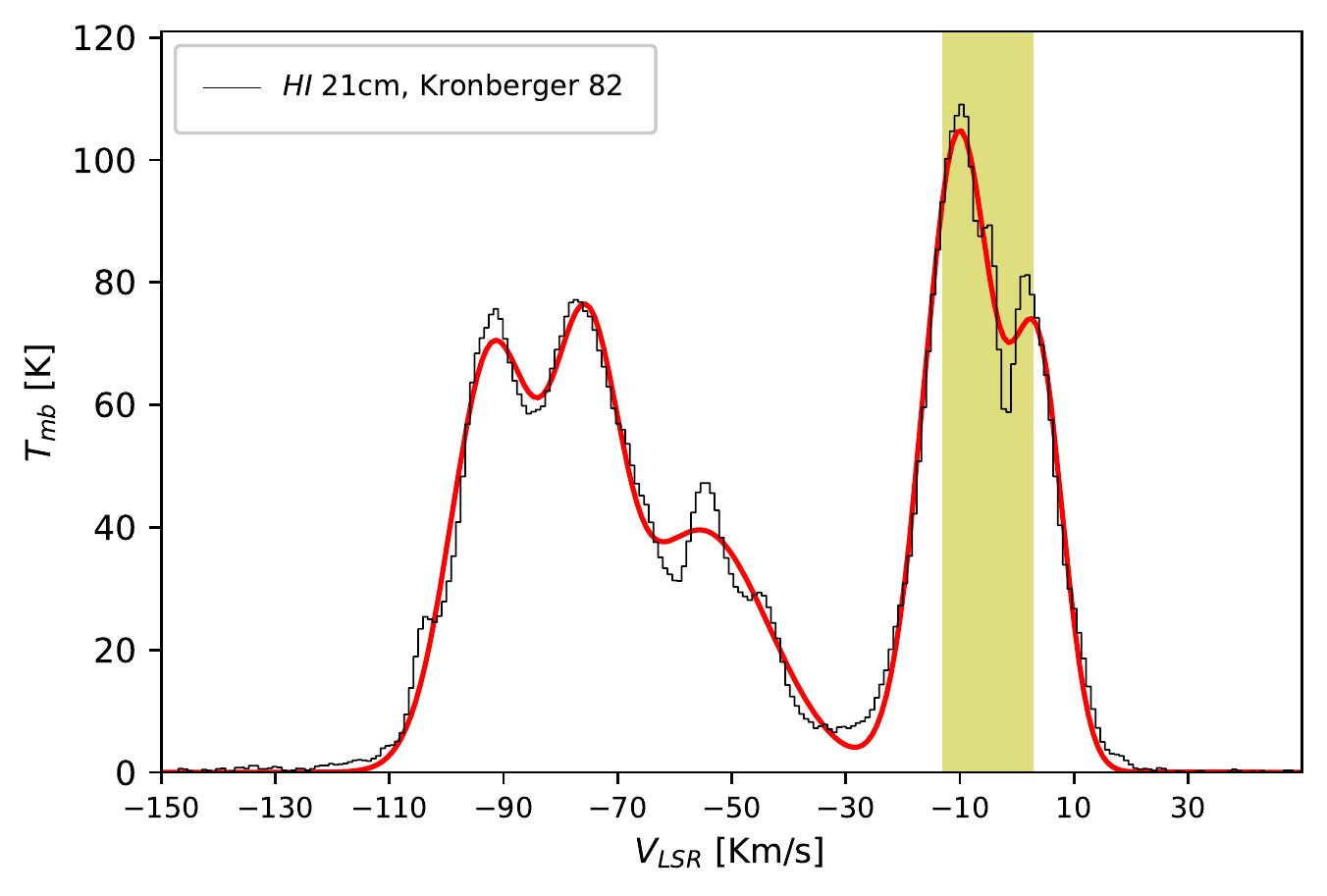}
\end{center}
\caption{Same as Fig. \ref{fig:spectra_2deg} but for Kronberger 82 (see Fig. \ref{fig:lhaaso_opt}). The weak component at $\sim$ 7 km s$^{-1}$ corresponds to COB-7 gas in the vicinity. 
 }
\label{fig:Spec_K82_CO}
\end{figure*}

In Fig.~\ref{fig:K82_COs} we show the low-resolution $^{12,13}$CO(J=2$\to$1) emission maps of the OPU around Kron 82 (cf. Fig. \ref{fig:CygOB7_intro} down ). The white square represents the region of $\approx$ 0.1 deg around the brightest emission where the spectra were extracted. The averaged spectra for $^{12,13}$CO(J=2$\to$1) and HI emissions are shown in Fig.~\ref{fig:Spec_K82_CO}. The fitting parameters are shown in Tab.~\ref{table:kron82_adj}. The black square, the crosses and the plus sign refer to objects studied by \citet{Moscadelli2021}. 

$^{12}$CO (J=2$\to$1) and $^{13}$CO (J=2$\to$1) spectra show the brightest principal component in a --13 $\gtrsim$ V$_{\rm LSR}$ $\gtrsim$ 3 km s$^{-1}$ range, centered at V$_{\rm LSR} \sim$ --6.6 km s$^{-1}$ (see Tab.~\ref{table:kron82_adj}). This value agrees with the systemic velocity of $\sim$ 6.1 km s$^{-1}$ reported by \citet{Moscadelli2021}. Besides the main line in the $^{12}$CO(J=2$\to$1) spectrum, we discover a weak wing component at $\sim$ --2.2 km s$^{-1}$. This component is probably related to the main velocity component of the CO gas in the COB7-MC region. On the other hand, the $^{12}$CO(J=2$\to$1) spectrum shows another weak component at V$_{\rm LSR}$ of 8.41 km s$^{-1}$ with an observed frequency of 230.5315 GHz and a signal-to-noise ratio $\sim$ 6.  We believe this line is related to CO molecular gas in COB7--MC, but high-resolution observations are needed to clarify its nature.

To estimate the physical parameters, we consider emission in the range V$_{\rm LSR} = $ -13 to 3 km s$^{-1}$ (see Fig. \ref{fig:Spec_K82_CO}), which includes both the main and wing components of the $^{12}$CO(J=2$\to$1) spectrum. The fitted parameters in this range are shown in Tab.~\ref{table:adj3}.

We follow the same methodology described in section \ref{sec:size} for the molecular cloud and use the peak brightness of the principal component to estimate the average excitation temperature using Eq.~\ref{eq:Tex} and the corresponding optical depth using Eq.~\ref{eq:tau}. We obtain an average of $T_{\rm ex} \sim$ 16 K within this region. The optical depth $\tau^{12}_{ CO }$ is then estimated from $\tau^{13}_{ CO }$. Next, the column densities of $^{13}$CO and $^{12}$CO are calculated using Eqs.~\ref{eq:13CO_column} and \ref{eq:12CO_column}, respectively. The H$_2$ column density is estimated using the equations ~\ref{eq:eqnNH2} and \ref{eq:12CO_column}. The results are shown in Tab. \ref{table:column3}. Finally, the number density of H$_2$ and H (H$_2$+ HI) is calculated using the distance to Kron 82 $\approx$ 1.63 kpc \citep{Moscadelli2021, Xu2013} and the Eq. \ref{eq:eqnnH}, which is shown in Tab.~\ref{table:column4}. The physical parameters agree with those of \citet{Moscadelli2021}.

\begin{figure*}

\includegraphics[width=0.49\textwidth]{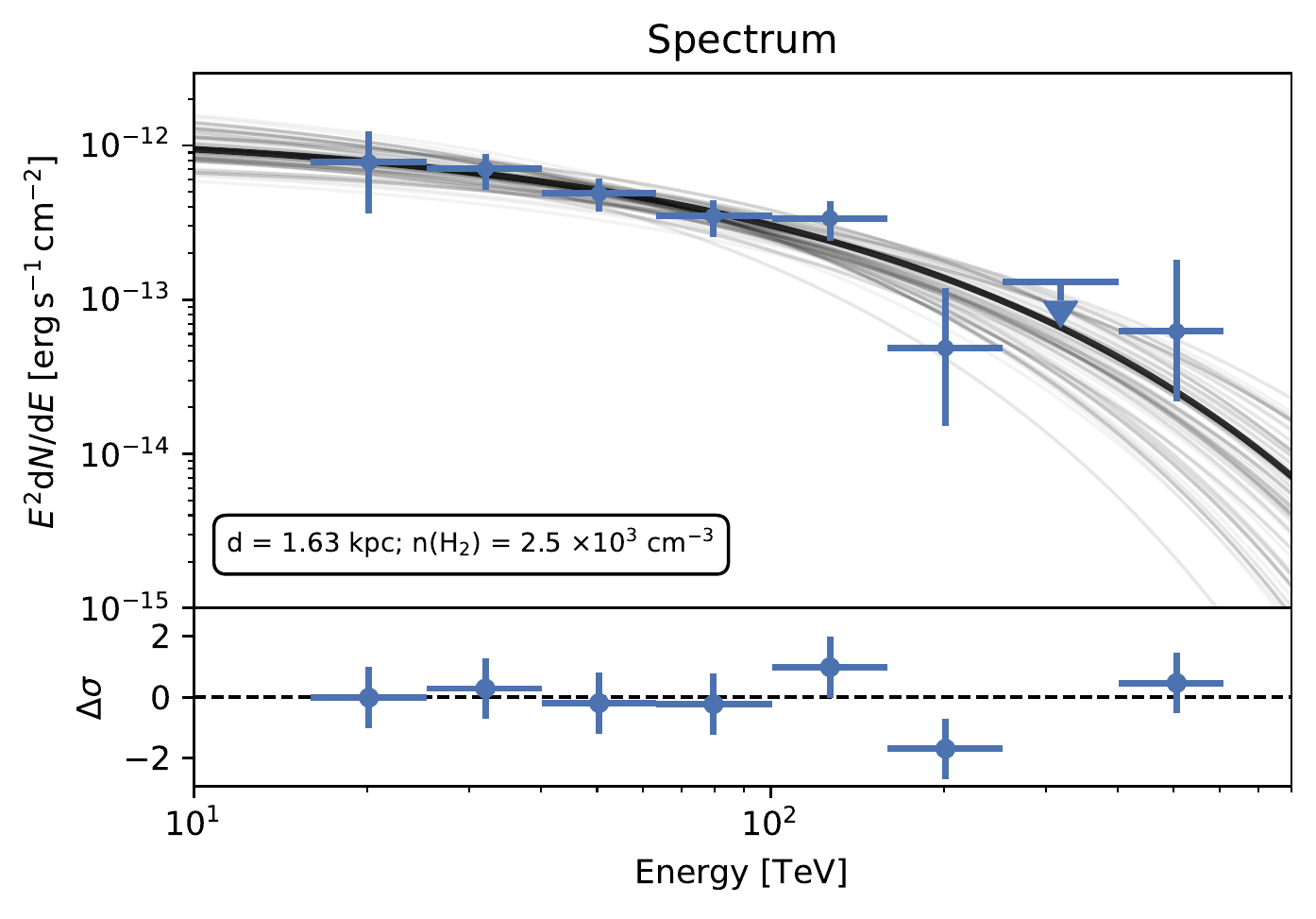}
\includegraphics[width=0.49\textwidth]{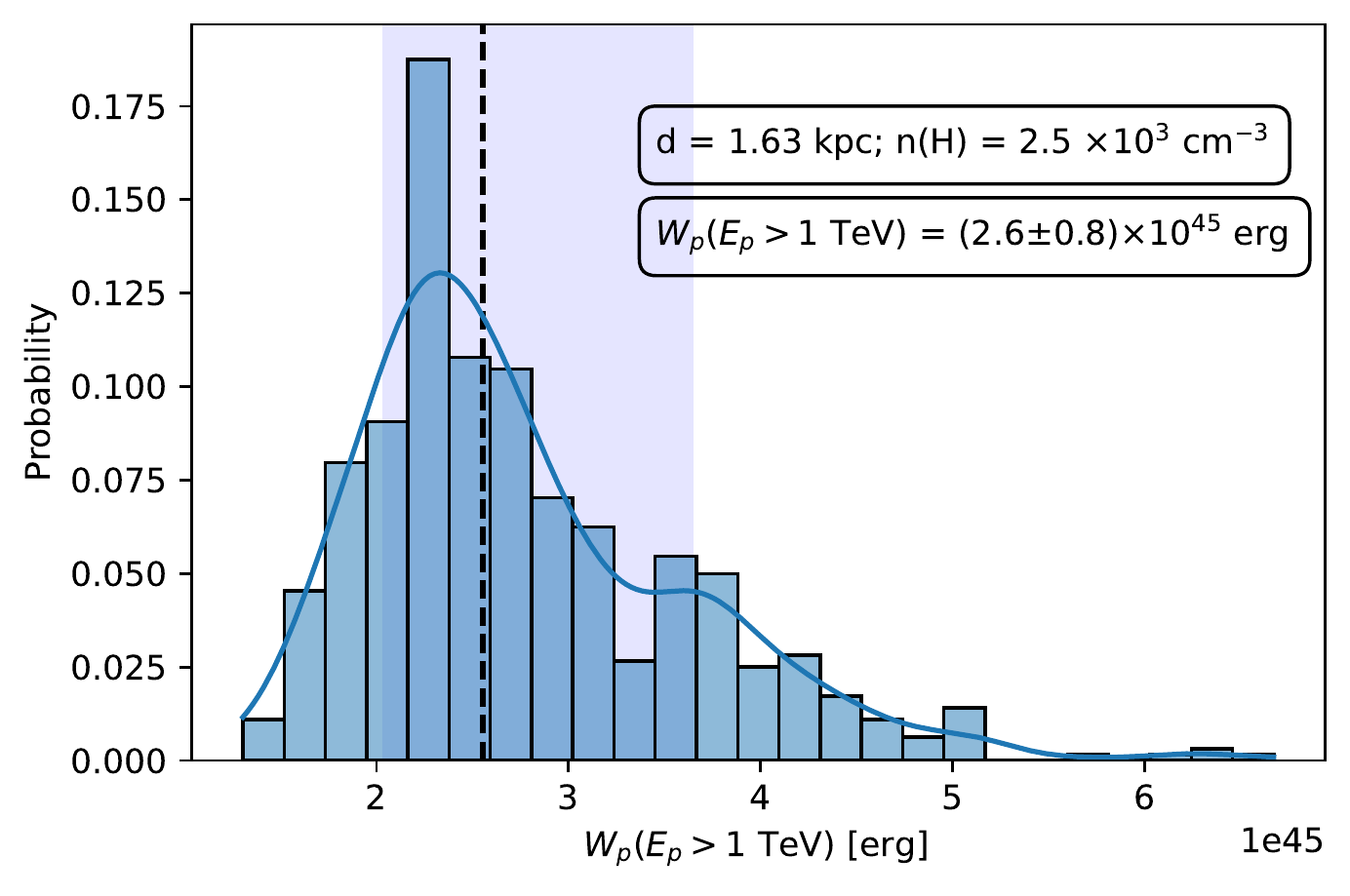}
\includegraphics[width=0.49\textwidth]{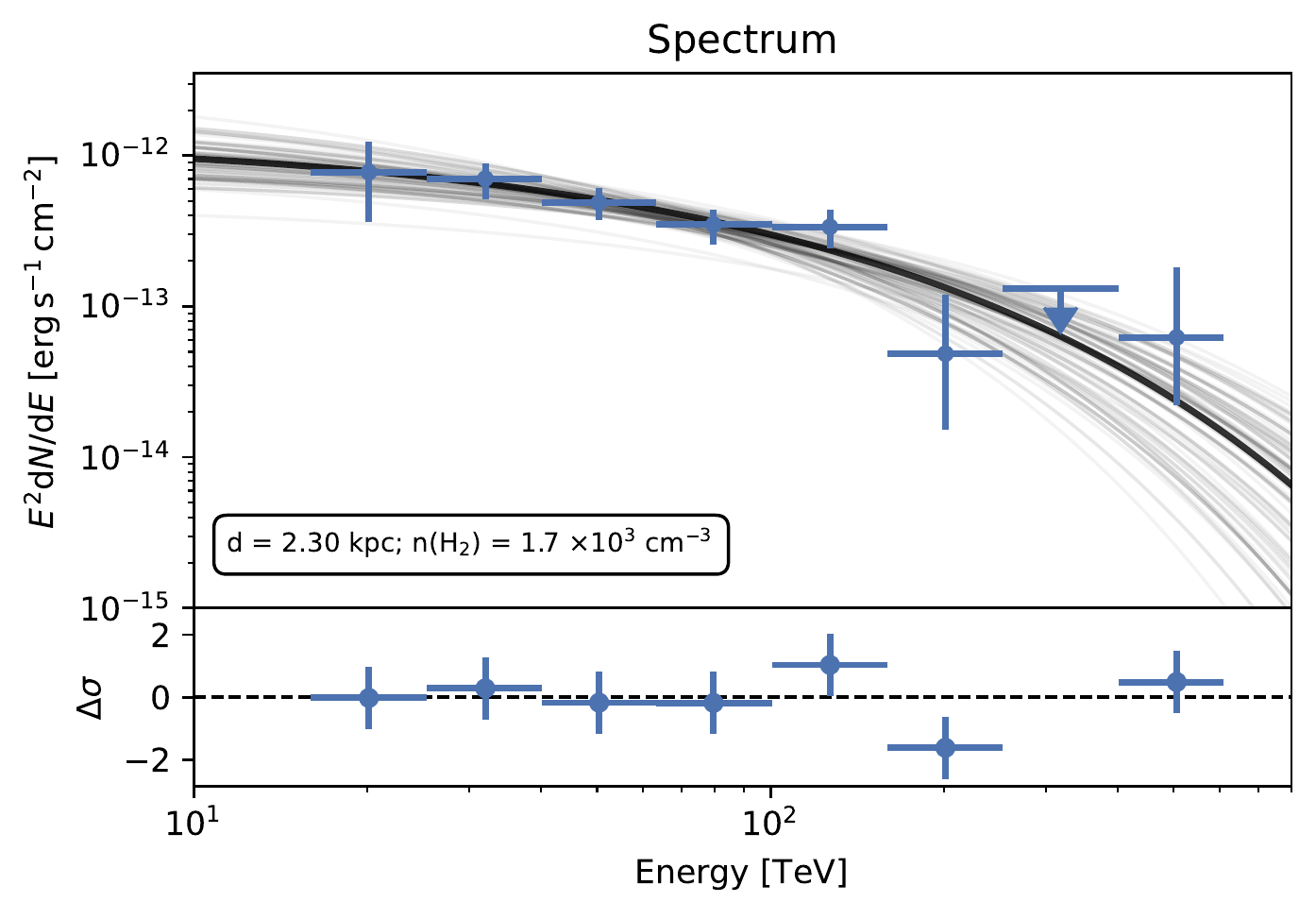}
\includegraphics[width=0.49\textwidth]{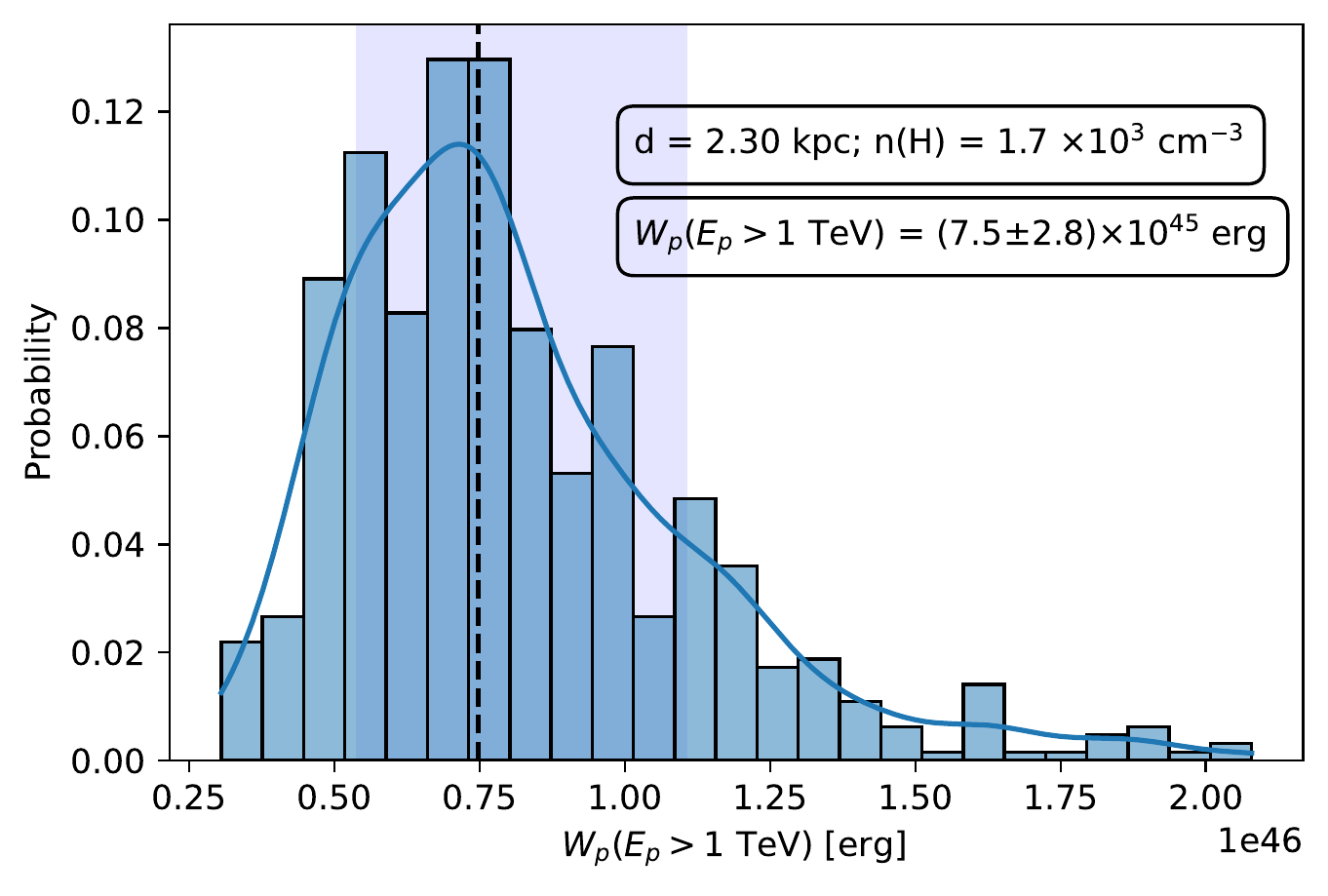}

    \caption{Same as Fig. \ref{fig:naima}, but for Kronberger 82. The hadronic model was calculated for the range of a) 1.63 kpc and b) 2.30 kpc. We use the respective nucleon density given in Tab. \ref{table:column4} and a cutoff of 600 TeV (see Tab. \ref{tab:naima_results_Kron82}). The total proton energy of $\sim$ 10$^{45}$ ergs required to reproduce the flux observed by LHAASO (blue points) is lower than those of [FKT-MC]2022 and MML (W$_p$ $\sim$ 10$^{47}$ ergs), so we cannot discard Kron 82 as a candidate for J2108 sub-PeV emission. }
    \label{fig:naima_K82}
\end{figure*}

On the other hand, using the rotation curve of \citep{Brand1993}, we obtain a far kinematic distance of 2.3$\pm$ 0.5 kpc. The parameters calculated at this distance are also shown in Tab.~\ref{table:column4} for comparison with those calculated at 1.63 kpc. The estimated physical parameters are similar at both distances. Thus, in a distance range between 1.63 and 2.30 kpc, Kron 82 is a molecular clump with a size of $\sim$ 0.1$^{\circ}$; a nucleon density (H$_2$+ HI) of about 10$^{3}$ cm$^{-3}$ and an M(H$_2$+ HI) of 10$^{3}$ M$_{\odot}$. The estimated H$_2$ number densities, H$_2$ total mass, and the viarl mass are shown in Tab.~\ref{table:mass_kron82}. We show the results for the computed hadronic model in Fig.~\ref{fig:naima_K82} and Tab.~\ref{tab:naima_results_Kron82}, taking into consideration these physical parameters and utilizing the same models derived by Naima as stated in ~\S~\ref{sec:gamma_res}. We calculated proton energies of 2.6 $\times$ 10$^{45}$ and 7.0 $\times$ 10$^{45}$ ergs for distances of 1.63 and 2.3 kpc, respectively.

Comparing these values to those of MML and [FKT-MC]2022 (9.0$\times$10$^{47}$ ergs and 1.6$\times$10$^{47}$ ergs; see Fig.~\ref{fig:naima}), Kron 82 requires significantly less cosmic-ray energy (proton) to replicate the emission of J2108, regardless of whether it was caused by a supernova explosion or massive stellar winds. These energies are $\sim$ 10$^{51}$ erg for supernova explosions and $\sim$ 10$^{52}$ erg for stellar winds from star clusters similar to Cygnus OB2 \citep[e.g.][and references therein] {Lozinskaya2002}. Even if Kron 82 appears to have a stellar component more similar to IRAS 21046+5110 than Cygnus OB2, it is possible for molecular gas to produce the LHAASO-measured energy for J2108. Nonetheless, due to its 0.4$^\circ$ radius separation from J2108, outside its (sub-PeV emission) upper limit radial extension \citep[$\sim$ 0.35$^{\circ}$;][]{Cao2021b}, this scenery is not favorable.

\begin{table*}
\caption{\centering Parameters of the adjusted Gaussian fits of the emissions of $^{12}$CO(J=2$\rightarrow$1) and $^{13}$CO(J=2$\rightarrow$1) around Kron 82. The main beam (MB) averaged peak temperature ($T^p_{\rm MB}$) uncertainties are only due to rms noise. Velocity channel resolutions are used to show the LSR velocity (V$_{\rm LSR}$) and FWHM ($\Delta V$) uncertainties.}
\centering
	\begin{tabular}{cccccc} 
		\hline
		Line & Diameter & V$_{\rm LSR}$ & $\Delta V$ & $T_{\rm MB}^{\rm peak}$ & $\int T_{\rm MB} dV$  \\
		& [deg] & $\rm km \ s^{-1}$ & $\rm km \ s^{-1}$ & [K] &[K km s$^{-1}$]\\
		\hline

$^{12}$CO 2-1 main &  0.10 $\pm$ 0.01 & -6.64 $\pm$ 0.08 & 3.511 $\pm$ 0.08 & 11.00 $\pm$ 0.40 & 41.10 $\pm$ 0.56 \\
$^{12}$CO 2-1 wing &  0.10 $\pm$ 0.01 & -2.24 $\pm$ 0.08 & 2.48 $\pm$ 0.14 & 2.46 $\pm$ 0.40 & 6.48 $\pm$ 0.45 \\
$^{12}$CO 2-1 weak & 0.10 $\pm$ 0.01 & 8.41 $\pm$ 0.08 & 2.44 $\pm$ 0.13 & 2.16 $\pm$ 0.40 & 5.60 $\pm$ 0.40 \\

$^{13}$CO 2-1 & 0.10 $\pm$ 0.01 & -6.39 $\pm$ 0.08 & 2.83 $\pm$ 0.08 & 4.30 $\pm$ 0.40 & 12.95 $\pm$ 0.41  \\

\hline
	\end{tabular}
	\label{table:kron82_adj}
\end{table*}

\begin{table*}
	\caption{\centering Values used for the column density analysis (Kronberger 82).}
	\centering
	\begin{tabular}{cccccc} 
		\hline
		Species & Diameter & $T_{\rm MB}^{\rm peak}$ & $\int T_{\rm MB} dV$ & $T_{\rm ex}$ & $\tau$  \\
		& [deg] & [K] &[K km s$^{-1}$] & [K] &    \\
		\hline

$^{12}$CO 2-1 main + wing & 0.10 $\pm$ 0.02 & 11.00 $\pm$ 0.40 & 47.58 $\pm$ 0.72 & 16.1 $\pm$ 0.4 & 29.2 $\pm$ 3.4 \\
$^{13}$CO 2-1 & 0.10 $\pm$ 0.02 & 4.30 $\pm$ 0.40 & 12.95 $\pm$ 0.41 & 16.1 $\pm$ 0.4 & 0.5 $\pm$ 0.1 \\
HI & 0.10 $\pm$ 0.02 & -- & 1379.3 $\pm$ 304.3 & -- & -- \\
\hline
	\end{tabular}
	\label{table:adj3}
\end{table*}

\begin{table*}
	\caption{\centering Estimated column (N) and number (n) densities of HI, $^{12}$CO, $^{13}$CO, and $\rm H_2$ for Kronberger 82. The 12 and 13 labels correspond to the respective CO emission used in the calculations.}
	\centering
	\begin{tabular}{cccccccc} 
		\hline
		Diameter & $N(\rm ^{13}CO)$ & $N(\rm ^{12}CO)$ & $N(\rm HI)$ & $N^{\rm ^{12}CO}(\rm H_2)$ & $N^{\rm ^{13}CO}(\rm H_2)$  & $n^{^{12}CO}(\rm H_2)$ & $n^{^{13}CO}(\rm H_2)$ \\
		deg & [$10^{15}$ cm$^{-2}$] & [$10^{17}$ cm$^{-2}$] & [$10^{21}$ cm$^{-2}$] & [$10^{21}$ cm$^{-2}$] & [$10^{21}$cm$^{-2}$] & [$10^{3}$cm$^{-3}$] & [$10^{3}$cm$^{-3}$] \\
		\hline

		0.10 $\pm$ 0.01 & 8.3 $\pm$ 0.3 & 6.8 $\pm$ 0.8 & 2.5 $\pm$ 0.5 & 9.5 $\pm$ 2.8 & 4.1 $\pm$ 1.2 & 1.1 $\pm$ 0.3 & 0.5 $\pm$ 0.1  \\

		\hline
    \end{tabular}
	\label{table:column3}
\end{table*}

\begin{table*}
	\caption{\centering Column (N) and number (n) densities of nucleons for Kronberger 82.}
	\centering
	\begin{tabular}{ccccccc} 
		\hline
		Diameter & Distance & $N^{12}(\rm H)$ & $N^{13}(\rm H)$ & $n^{12}(\rm H)$ & $n^{13}(\rm H)$  \\
		$[\rm deg]$ & [kpc] & [$10^{22} \rm cm^{-2}$] & [$10^{22} \rm cm^{-2}$] & [$10^3 \rm cm^{-3}$] & [$10^3 \rm cm^{-3}$]   \\
		\hline

0.10 $\pm$ 0.01 & 1.63 $\pm$ 0.05 & 1.1 $\pm$ 0.3 & 2.2 $\pm$ 0.6 & 2.5 $\pm$ 0.7 & 1.2 $\pm$ 0.3 \\
0.10 $\pm$ 0.01 & 2.30 $\pm$ 0.50 & 1.1 $\pm$ 0.3 & 2.2 $\pm$ 0.6 & 1.7 $\pm$ 0.6 & 0.9 $\pm$ 0.3 \\

\hline
	\end{tabular}
	\label{table:column4}
\end{table*}

\begin{table*}
	\caption{\centering Parameters of the molecular emission for Kronberger 82}
	\centering
	\begin{tabular}{ccccccc} 
		\hline
		Source & V$_{\rm LSR}$  & $D$ & Distance & $M_{\rm vir}({\rm H_2})$ & $M({\rm H_2})$  & $M(\rm HI + H_2)$ \\
		& [$\rm km \ s^{-1}$]  & [deg] & [kpc] & [$10^4 M_\odot$] & [$10^3 M_\odot$] & [$10^3 M_\odot$] \\
\hline
Kronberger 82 ($^{13}$CO) & -6.6 $\pm$ 0.1 & 0.10 $\pm$ 0.01 & 1.63 $\pm$ 0.05 & 2.4 $\pm$ 0.1  & 0.4 $\pm$ 0.1 & 0.5 $\pm$ 0.1 \\
Kronberger 82 ($^{12}$CO) & -6.6 $\pm$ 0.1  & 0.12 $\pm$ 0.01 & 1.63 $\pm$ 0.05 & 3.7 $\pm$ 0.1 & 1.5 $\pm$ 0.5 & 1.7 $\pm$ 0.5\\
Kronberger 82 ($^{13}$CO) & -6.6 $\pm$ 0.1 & 0.10 $\pm$ 0.01 & 2.33 $\pm$ 0.51 & 2.4 $\pm$ 0.1  & 0.3 $\pm$ 0.1 & 0.4 $\pm$ 0.1 \\
Kronberger 82 ($^{12}$CO) & -6.6 $\pm$ 0.1  & 0.12 $\pm$ 0.01 & 2.33 $\pm$ 0.51 & 3.7 $\pm$ 0.1 & 1.1 $\pm$ 0.4 & 1.2 $\pm$ 0.4\\
\hline

	\end{tabular}
	\label{table:mass_kron82}
\end{table*}

\

\begin{table*}
    \caption{\centering Naima Results for Kronberger 82}
    \centering
    \begin{tabular}{ccccccc} \hline
        & Distance & n(\rm{H}) & ROI & $W_p$ & Cutoff & Obtained \\ 
         & [kpc] & [cm$^{-3}$] & [deg] & [$10^{45}$erg] & [TeV] & from \\ \hline
    Kron 82 2022 & 1.63 & 2.5 $\times$ 10$^3$ & 0.1 & 2.6$_{-0.5}^{+1.1}$ & $600 \pm 300$ & $^{13}$CO(J=2$\rightarrow$1) \\
      Kron 82 2022 & 2.30 & 1.7 $\times$ 10$^3$ & 0.1 & 7.0$_{-2.0}^{+4.0}$ & $600_{-200}^{+400}$ & $^{13}$CO(J=2$\rightarrow$1)
    \end{tabular}
    \label{tab:naima_results_Kron82}
\end{table*}

\end{document}